# Correlation of Fe-based Superconductivity and Electron-Phonon Coupling in an FeAs/Oxide Heterostructure

Seokhwan Choi[1], Steven Johnston[2], Won-Jun Jang[1,3†], Klaus Koepernik[4], Ken Nakatsukasa[2], Jong Mok Ok[5], Hyun-Jung Lee[5], Hyun Woo Choi[1], Alex Taekyung Lee[6], Alireza Akbari[5,7], Yannis K. Semertzidis[1,3], Yunkyu Bang[8], Jun Sung Kim[5], Jhinhwan Lee[1*]

*[1] Department of Physics, Korea Advanced Institute of Science and Technology, Daejeon 34141, Korea*
*[2] Department of Physics and Astronomy, University of Tennessee, Knoxville, Tennessee 37996, USA*
*[3] Center for Axion and Precision Physics Research, Institute for Basic Science (IBS), Daejeon 34051, Republic of Korea*
*[4] IFW Dresden, P.O. Box 270116 Dresden, Germany*
*[5] Department of Physics, Pohang University of Science and Technology, Pohang 37673, Korea*
*[6] Department of Applied Physics and Applied Mathematics, Columbia University, New York 10027, USA*
*[7] Asia Pacific Center for Theoretical Physics, Pohang, 790-784, Korea*
*[8] Department of Physics, Chonnam National University, Kwangju 500-757, Korea*

*† Current affiliation: Center for Quantum Nanoscience, Institute for Basic Science (IBS), Seoul 03760, Republic of Korea; Department of Physics, Ewha Womans University, Seoul 03760, Republic of Korea*

*e-mail: jhinhwan@kaist.ac.kr

(Dated: July 31, 2017)

**Abstract**

Interfacial phonons between iron-based superconductors (FeSCs) and perovskite substrates have received considerable attention due to the possibility of enhancing preexisting superconductivity. Using scanning tunneling spectroscopy, we studied the correlation between superconductivity and e-ph interaction with interfacial-phonons in an iron-based superconductor $Sr_2VO_3FeAs$ ($T_c \approx 33$ K) made of alternating FeSC and oxide layers. The quasiparticle interference measurement over regions with systematically different average superconducting gaps due to the e-ph coupling locally modulated by O vacancies in $VO_2$ layer, and supporting self-consistent momentum-dependent Eliashberg calculations provide a unique real-space evidence of the forward-scattering interfacial phonon contribution to the total superconducting pairing.

The discovery of high temperature superconductivity in monolayer iron-based superconductors (FeSC) on perovskite substrates [1-7] has attracted broad attention due to the possibility of achieving $T_c$ in the FeSCs comparable to the cuprate superconductors. Importantly, these interface systems suggest an entirely new mechanism for enhancing superconductivity, applicable to a broad range of layered superconducting heterostructures: forward-scattering interfacial phonons [2,4,6,7]. This principle is evidenced by an angle-resolved photoemission spectroscopy (ARPES) study [2], which revealed that the increased $T_c$ in $FeSe/SrTiO_3$ coincides with the appearance of the replica bands as the thickness of the FeSe layers is varied. Forward scattering by interfacial phonons is believed to produce these replicas, and this interaction has been proposed as a general means to enhance $T_c$ in heterostructures [6]. However, i) a definitive example of this principle in a bulk heterostructure suitable for practical application and ii) a direct real-space evidence of interfacial-phonon enhanced superconductivity are currently lacking.

To solve these issues, a bulk heterostructure made of alternating FeSC monolayers and perovskite layers has been suggested [4,6]. Interestingly, an iron-based superconductor $Sr_2VO_3FeAs$ with highest $T_c \approx 33$ K among all the 21311-family [8-11] has a structure nearly equivalent to the suggestions and therefore is a unique *bulk* candidate suitable for investigating the effect of interfacial-phonons on superconductivity. $Sr_2VO_3FeAs$ also exhibits self-doping by interlayer charge transfer between the perovskite $Sr_2VO_3$ layers and the FeAs layer [12], and shows a preferred and reproducible symmetric cleavage at the SrO-SrO interface. These properties make it an ideal system for surface spectroscopic study using a real-space probe, i.e. scanning tunneling microscopy and spectroscopy (STM/STS). STM/STS has been used to map the spatial inhomogeneity of superconducting gap in many unconventional superconductors, and the correlations between the gap and the locations of dopant atoms, magnetic vortices or the supermodulation of lattices have been widely studied [13-19].

In this letter, we report QPI measurements on the bulk heterostructured superconductor $Sr_2VO_3FeAs$ grown using self-flux technique [10]. We observe both the filled and empty state replica bands in the QPI data, which is a spectroscopic evidence of e-ph coupling with forward-scattering phonons. We notice various changes in the renormalized bands in regions with systematically different superconducting gaps due to locally varying e-ph coupling near particular defects. These would constitute a unique demonstration of enhancement of superconductivity by forward-scattering phonons in a bulk heterostructured Fe-based superconductor system where a peak-hump analysis is not allowed.

Figure 1(a) shows a SrO-terminated topograph with randomly oriented $C_2$ domains. These domains are attributed to surface reconstructions since all bulk measurements consistently reported no orthorhombic distortion from room temperature down to 4.6 K [9-11]. The temperature dependence of a typical *dI/dV* spectrum is shown as the inset of Fig. 1(c). The normal state spectrum at $T$ = 140 K displays an asymmetric shape common to many FeSCs [20,21]. At lower temperatures, we further observe the opening of a spectroscopic gap at $T \sim$ 30 K, which is evident in both the temperature dependent averaged spectra (Fig. 1(d)) and the QPI intensity (Figs. 2(c) and S2) near the Fermi level ($E_F$). Overall it is consistent with the $T_c$ observed in bulk magnetic susceptibility measurements (inset of Fig. 1(a)). The electron density plot obtained from band structure calculations integrated near the Fermi level shows that the electron tunneling of STM primarily probes the FeAs layer for bias voltages near $E_F$ and is thus sensitive to the bulk FeAs bands [22]. We therefore ascribe the gap near $E_F$ to the bulk superconducting gap. We also observe an additional temperature dependent peak around -18 meV and the finite zero-bias *dI/dV* not often found in other FeSCs with simpler structures [35-38]. At this time their origins are not clear; however, they are not significant in reaching the main conclusion of this letter. At least they are virtually independent of the topmost SrO layer surface reconstruction (See SM section 10, Ref. [39] and Ref. [40]). The -18 meV peak may be related to i) the hole-like $\beta$ band located below the $\alpha$ band (resulting in blue dashed curve in Fig. 2(c) with $m^* \approx 4.2 m_e$), ii) spin-density-wave (SDW) gap edges associated with the Fe magnetic order experimentally observed in $^{75}$As NMR experiment below 45 K [10,20], and/or iii) many body effects related to doping the strongly correlated V-derived Hubbard bands [22].

To image the renormalized band structure for both filled and empty states with sufficiently high resolution suitable for mask-based QPI analysis, we performed a spectroscopic imaging STM (SI-STM) measurements with higher spatial and energy resolutions than typically required: Figure 2(c) shows a Fourier-transformed ($q$-space) image of QPI based on a (300 nm)$^2$, 512×512 pixel $dI/dV$ map taken over an energy range [–40 meV, +40 meV] with 201 energy layers. There we see a dominant parabolic band that crosses $E_F$ with a band minimum near –12 meV, consistent with the shallow electron-like Fe $d_{xz}/d_{yz}$-derived $\alpha$ band at Γ point observed marginally by ARPES in $Sr_2VO_3FeAs$ [41,42] and clearly in other FeSC under certain doping conditions [43,44]. From its intraband QPI dispersion (denoted $\alpha$-$\alpha$) in Fig. 2(c), we extract a light effective mass $m^* \approx 0.47 m_e$, implying a large in-plane overlap of the Fe orbitals making up this band. The possibility that this band originates from the M electron pocket is ruled out because of its small pocket size and light effective mass compared with the M bands observed in ARPES [41-44]. The superconducting gap observed in our *dI/dV* spectra is visible as reduced intensity at $E_F$ in the $\alpha$-$\alpha$ QPI dispersion (Fig. 2(c)). The gap is also weakened by application of a magnetic field and suppressed at temperatures above the bulk $T_c$ (Fig. S2), which show that it is related to bulk superconductivity and that the $\alpha$ band electrons participate in superconducting pairing.

The signatures of e-ph coupling are clear in the QPI data, as shown in Fig. 2(c). For example, the filled-state portion of the $\alpha$-$\alpha$ QPI dispersion has a replica band ($\alpha''$-$\alpha''$) shifted down by about 14 meV ($\approx$ phonon energy $\Omega_{ph}$), which persists well above $T_c$ (see Fig. S2). This is reminiscent of the replica bands of the M electron pockets persisting well above $T_c$ in ARPES experiments on the $FeSe/SrTiO_3$ system [2,7,45,46]. The empty-state portion of the $\alpha$-$\alpha$ band also has a replica band ($\alpha'$-$\alpha'$) shifted up by $\sim\Omega_{ph}$. Self-consistent high-order e-ph calculations [45] explain these replica features reasonably well by considering an e-ph coupling to a forward scattering optical phonons with $\Omega_{ph} \approx$ 14 meV and $g_{ph} \approx \Omega_{ph}$, where $g_{ph}$ is the strength of the e-ph coupling.

To gain further insight into these replica features, we performed an LDA-calculation of the phonon dispersion in $Sr_2VO_3FeAs$ and calculated the e-ph coupling strength with a frozen phonon calculation [22]. The results show that the phonon mode responsible for the replica bands is the symmetric breathing $c$-axis optical mode sketched in Fig. 2(g). It has an energy of 13.1 meV at the Γ-point and its coupling is strongly forward focused with $g_{ph} \approx$ 11 meV as shown in Figs. 2(f) and S8(c).

Using these estimates, we simulated the spectral function (SF, $A(\boldsymbol{k}, w)$) based on self-consistent Migdal-Eliashberg theory and the QPI using self-consistent T-matrix formalism [45]. Figures 2(d) and 2(e) reproduce the calculated SF and QPI maps, respectively. In the weak coupling and perfect forward scattering limits, the e-ph interaction produces filled and empty states replica bands at energies shifted by $\sim\Omega_{ph}(1+\lambda)$ with spectral weight proportional to $\sim\lambda$, where $\lambda \equiv g_{ph}^2/\Omega_{ph}^2$ is the dimensionless coupling constant [7,22]. The empty state dispersion of the original band shifts down and the filled state dispersion of the original band shifts up by similar amounts proportional to $\sim\lambda$.

To reveal the correlations between the superconducting gap $\Delta$ and the e-ph coupling constant $\lambda$, we performed a mask-based QPI analysis taking advantage of the spatial inhomogeneity of the $\Delta$ map due to random distribution of particular defects, as shown in Figs. 1(b) and 4(b). This novel analysis method is the only choice where the more conventional analysis [16-19] of point-by-point correlation of the coherence peak and the bosonic satellite hump is prohibited due to the strong SDW-gap-edge-like features located near the hump energy (near $\pm 18$ meV). We first divide the field-of-view into two regions (denoted $\Delta^{+}$ and $\Delta^{-}$) intertwined in coherence length scale based on the superconducting gap map. The $\Delta^{+}$ ($\Delta^{-}$) region corresponds to the area with $\Delta$ larger (smaller) than the median gap value so that they are identical in area and their Fourier-transformed intensities can be directly compared. The QPI modulations in each of these regions are then independently Fourier-transformed and plotted as a function of $\boldsymbol{q}$ and $V$, as shown in Figs. 3(a), 3(b) and S3-S6 [22]. This method allows us to systematically correlate the superconducting pairing strength $\Delta$ to various subtle features of the band renormalization that are dependent on e-ph coupling strength $\lambda$ without introducing mask-dependent artifacts (Ref. Fig. S4).

To visualize the detailed changes in band renormalization associated with changes in $\lambda$ possibly correlated with $\Delta$, we show contour plots of QPI data corresponding to $\Delta^{+}$and $\Delta^{-}$ regions, as shown in Fig. 3(e). The red ($\Delta^{+}$) and blue ($\Delta^{-}$) contours clearly show the shifts of the original $\alpha$-$\alpha$ band segments toward $E_F$. The details of the trend are all consistent with band renormalization with increased $\lambda$ [22] that are reproduced in self-consistent QPI simulations for two $\lambda$ values, which reproduces the experimental data as shown in Figs. 3(c),(d), and (f). The agreement, together with the conclusions from the next two paragraphs, provides atomic-scale spectroscopic evidence of an e-ph interaction positively correlated with the superconducting gap. The self-consistently calculated gap values (~4 meV) are about half of the experimental gap values (~7 meV) [22], showing that a significant fraction of the total superconducting pairing comes out of e-ph coupling.

To explain the source of the gap inhomogeneity, we focused on the peanut-shaped $C_2$ defects that appear randomly scattered in the *dI/dV* image near -18 meV as shown in Fig. 4(a). We do not observe any unitary in-gap bound states in the vicinity of this impurity as shown in the inset of Fig. 4(c), which rules out defects in the FeAs plane [22,35,47,48]. The only remaining possibility considering its $C_2$ symmetry is that it corresponds to a missing O ion in the $VO_2$ layer. Its correlation with the superconducting gap is significantly high as shown in Fig. 4(b) and 4(c), considering the fact that e-ph interaction is only an additional pairing mechanism here and that we can only see half of all the O vacancies neighboring an FeAs layer (i.e. the O vacancies in the upper $VO_2$ layer above the top most FeAs layer) due to the tunneling geometry. The dominant contribution to the Coulomb potential oscillation for the symmetric breathing phonon mode comes from the positively charged Sr ions closest to the FeAs plane, and all the ions in one perovskite layer on one side of FeAs layer are moving in phase for this mode. Therefore, the absence of a negatively charged O ion at each $C_2$ vacancy site would result in less cancellation of the Coulomb potential oscillation effect of the closest Sr ions, making the e-ph coupling significantly higher, i.e. about 17 % increase of $\lambda$ based on the frozen phonon calculation (See SM section 7). The measured gap size increase of 0.2 meV in the presence of the O vacancies can be compared with the theoretically calculated upper limit of 0.33 meV based on the Eqs. S7.7-S7.9. Furthermore, from the fact that in the higher $\Delta$ region the filled and empty state dispersions of the α band move towards the Fermi level as shown in Fig. 3(e), rather than moving together downward as can be expected for electron-doping, we can rule out the possibility of charge-transfer-induced superconducting gap increase. We can also exclude the possibility that the replica bands could be due to tunneling into and out of virtual states coupled with real states via emission of forward scattering phonons (See SM section 2).

In summary, we performed QPI measurements on a bulk FeAs/oxide heterostructure superconductor $Sr_2VO_3FeAs$. We observed the filled and empty state replica bands in the QPI data as a spectroscopic evidence of forward-scattering phonons of energy $\Omega_{ph} \approx 13.1$ meV corresponding to a symmetric breathing $c$-axis optical phonon mode. Using gap-map-based masked QPI analysis and self-consistent QPI calculations, we observed a number of subtle changes in the renormalized bands consistent with the positive correlation between the e-ph coupling and the superconducting gap in coherence length scale. The source of the gap inhomogeneity could be explained not by charge doping effect but by change in local e-ph coupling strength in the presence of O vacancies in the $VO_2$ layer for this phonon mode. Our results provide a unique demonstration of phonon-induced enhancement of unconventional superconductivity in a bulk heterostructure system comprised of FeAs/oxide layers.

**Acknowledgement**

This work was supported by the Metrology Research Center Program funded by Korea Research Institute of Standards and Science (No. 2015-15011069), the Pioneer Research Center Program through the National Research Foundation (NRF) (No. 2013M3C1A3064455), the Basic Science Research Programs through the NRF (No. 2013R1A1A2010897 and No. 2012R1A1A2045919), the Institute of Basic Science (IBS-R017-D1), the Brain Korea 21 PLUS Project of Korea Government, the NRF through SRC Center for Topological Matter (No. 2011-0030046), the Max Planck POSTECH/KOREA Research Initiative Program (No. 2011-0031558) through NRF funded by MSIP of Korea, the NRF through MSIP of Korea (2015R1C1A1A01052411). AA is grateful to the Max Planck Institute for the Physics of Complex Systems (MPI-PKS) for the use of computer facilities.

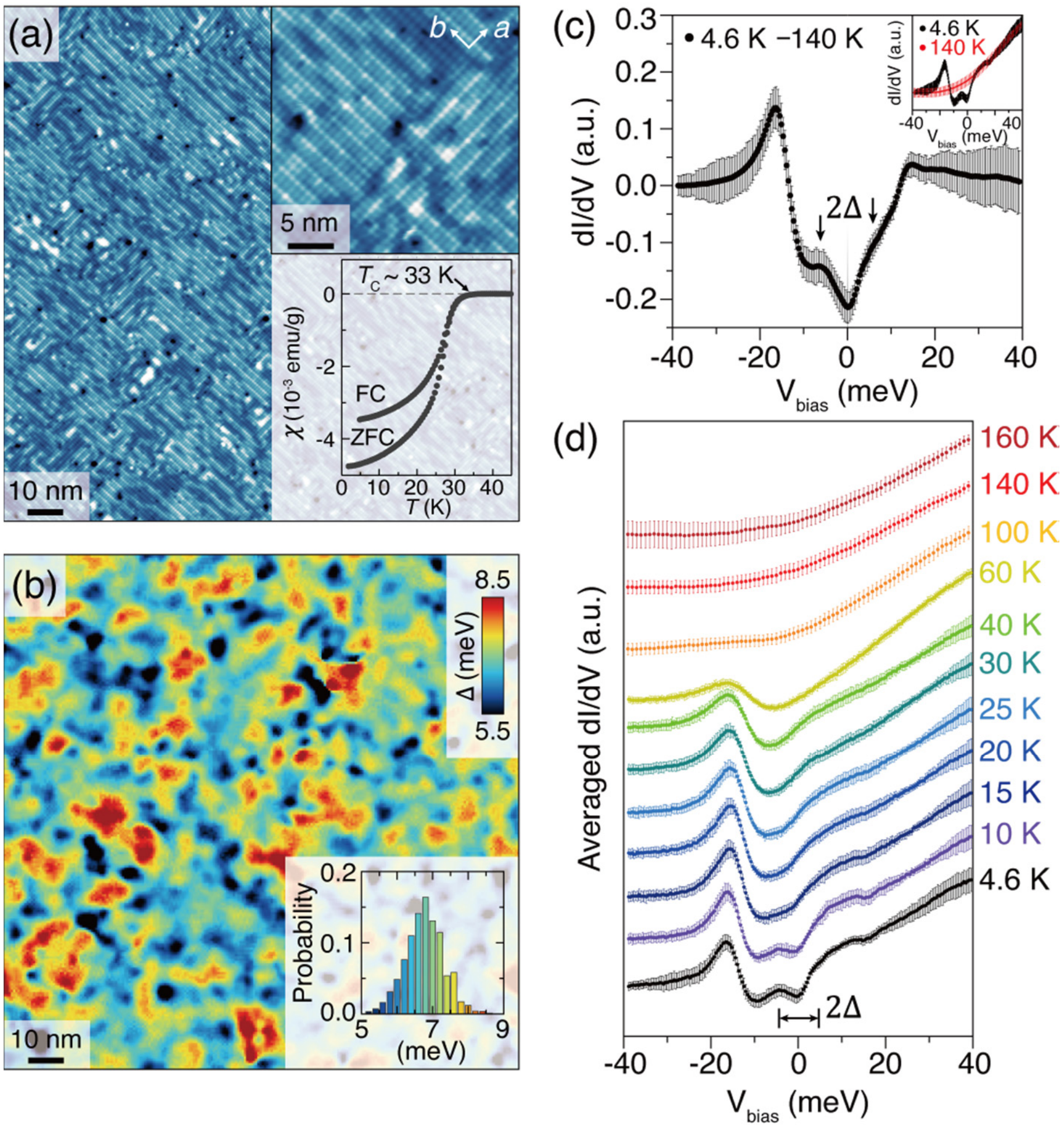

FIG. 1. (a) Topograph of the SrO-terminated surface of $Sr_2VO_3FeAs$ showing quasi-$C_2$-symmetric surface reconstructions. The lower right inset shows the magnetic susceptibility vs. temperature, showing bulk superconductivity near 33 K. (b) Superconducting gap ($\Delta$) map taken over the same area shown in the topograph of (a). (c) Averaged spectra (inset) taken at 4.6 K and 140 K and their difference showing the superconducting gap near the Fermi level marked with arrows. (d) Temperature-dependence of spatially averaged tunneling spectra. The error bars in (c) and (d) correspond to 0.5 times the standard deviation taken over 256 different locations.

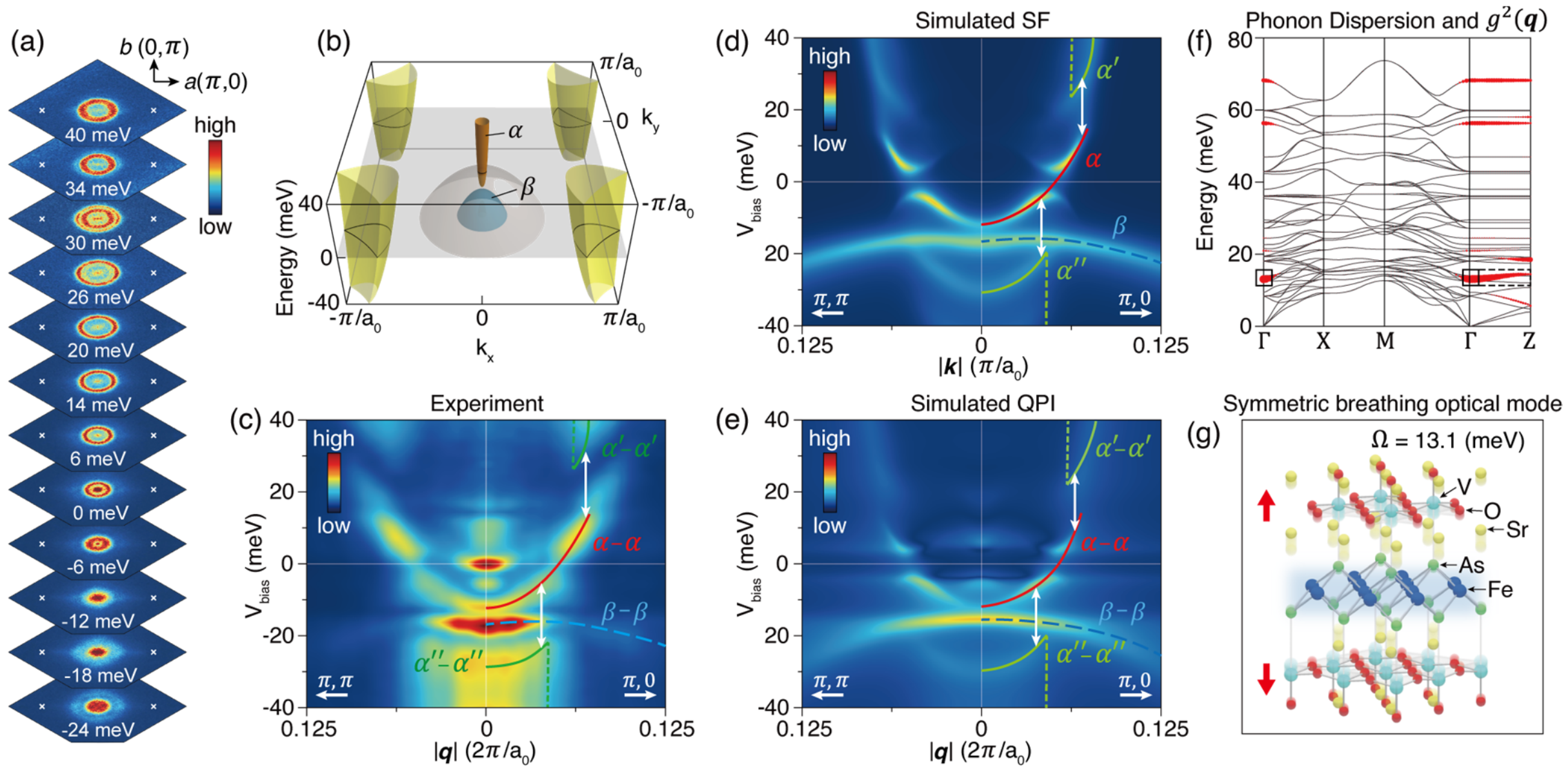

FIG. 2. (a) Quasiparticle interference maps $dI/dV(\boldsymbol{q},V)$, shown at various representative energies. The crosses mark $\boldsymbol{q} = \left(\pm\frac{1}{4}\frac{2\pi}{a_0},0\right)$. (b) The band structure (two-iron unit cell) of $Sr_2VO_3FeAs$ near the Fermi level, based on ARPES measurements and this work. The isotropic paraboloid (orange) is the Fe $d_{xz}/d_{yz}$ band (denoted $\alpha$ band) at $\Gamma$ point seen in Fig. S1(b,c). (c) A cross-section of the 3-dimensional quasiparticle interference $dI/dV(\boldsymbol{q},V)$ along two different crystallographic orientations $(\pi,\pi)$ and $(\pi,0)$. (d),(e) A cross-section of the simulated spectral function (SF) (d) and QPI (e) with $\lambda \equiv g_{ph}^2/\Omega_{ph}^2 = 0.69$ [22]. The curves marked $\alpha'$ and $\alpha''$ are the replicas of the $\alpha$ band with apparent energy shifts by $\sim\Omega_{ph}$. (f) The phonon dispersion in $Sr_2VO_3FeAs$, where the fat band thickness indicates the Coulomb coupling strength of each mode to the Fe layer electrons. The black boxes represent the symmetric breathing phonon mode most strongly coupled with the itinerant Fe electrons, whose atomic displacement snapshot is shown in (g) [22].

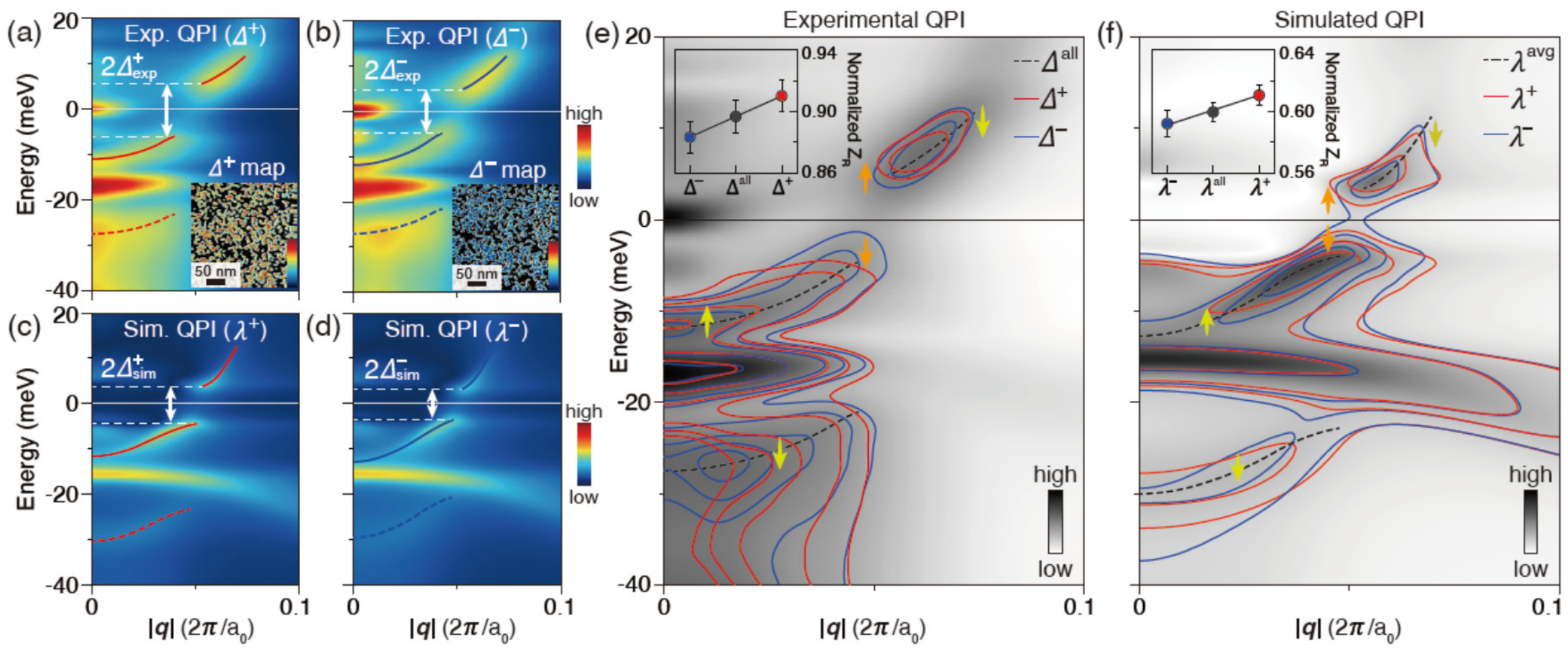

FIG. 3. (a),(b) Experimental *dI/dV*(|***q***|,V) evaluated over two intertwined areas determined by superconducting gap higher ($\Delta^+$) (a) and lower ($\Delta^-$) (b) than the median gap value. The insets show corresponding masks applied to the original gap map (300 nm × 300 nm). (c),(d) Images corresponding to (a),(b) for the simulated QPI intensity with parameters $\lambda^+$= 0.75 and $\lambda^-$= 0.63. The red and blue solid (dashed) curves in (a)-(d) indicate the EDC maxima of original (replica) bands. (e) The contour plots of both (a) and (b) with the background showing the unmasked original *dI/dV*(|***q***|,V). (f) The contour plots of both (c) and (d) with the background showing their sum. The black dashed lines in (e) and (f) indicate the EDC maxima of the backgrounds. As shown in (e), (f) and their insets, the increase of the replica band intensity $Z_R$, the band shifts of the filled and the empty state $\alpha$ band segments toward the Fermi level (yellow arrows) and the down-shift of the filled-state replica band segment (yellow arrows) are visible as the superconducting gap is increased (orange arrows).

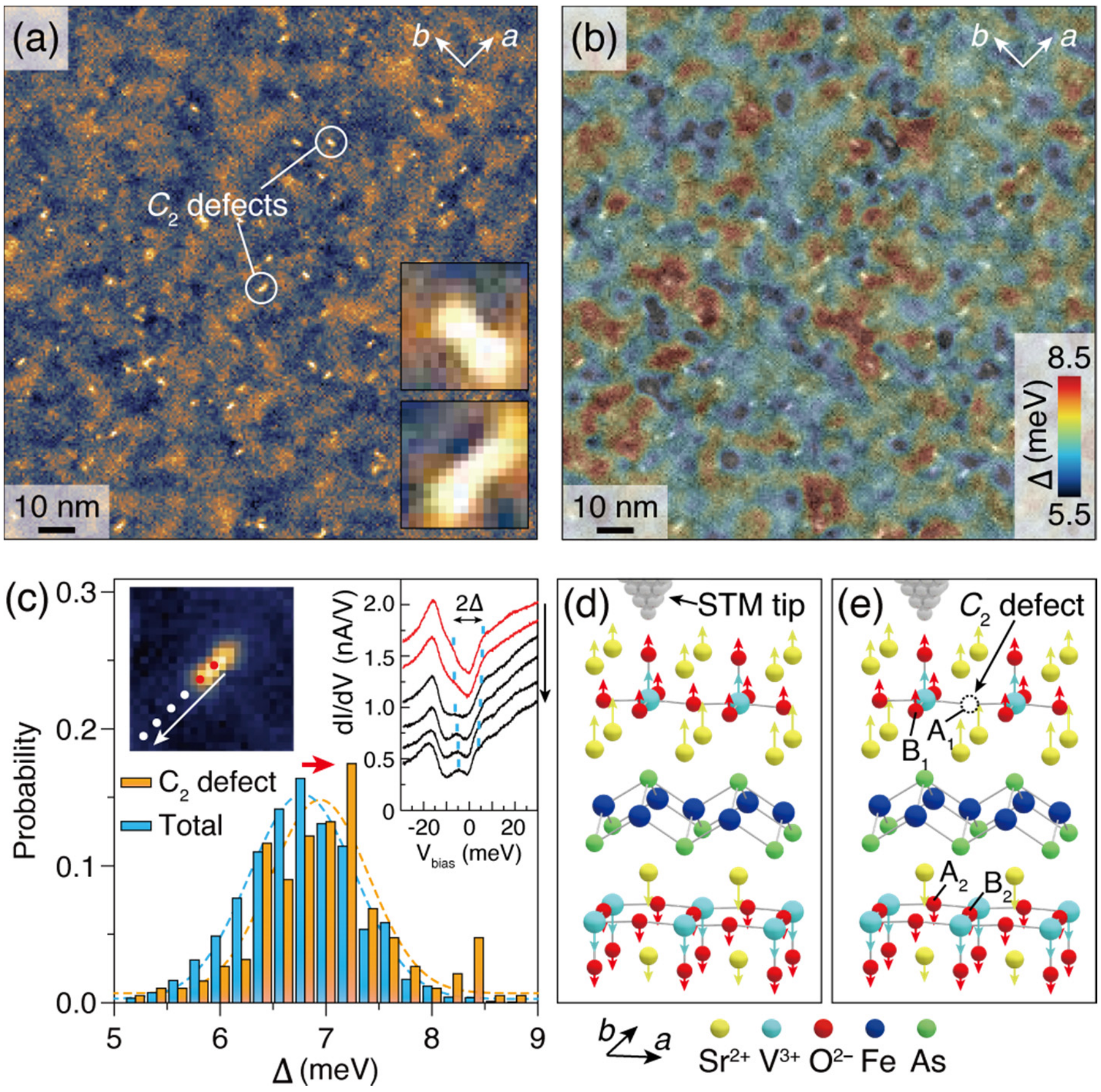

FIG. 4. (a) *dI/dV* map at -18 meV showing the $C_2$ defects corresponding to O vacancies in the $VO_2$ layer above the topmost FeAs plane. (b) The *dI/dV* map in (a) overlaid on the gap map in Fig. 1(b). (c) Gap histograms taken over all field of view and over $C_2$ defects only, with their fitted Gaussian curves shown. The right inset shows *dI/dV* spectra taken over the six points on (red) and away from (black) a $C_2$ defect shown in the left inset. (d),(e) Crystal structures without and with a $C_2$ defect induced by an O vacancy in $VO_2$ layer. Arrows indicate the proportionally exaggerated ionic displacements for the symmetric breathing phonon mode. The O vacancies at $A_2$ and $B_2$ sites are invisible to our tunneling measurement but they should exist and affect the superconductivity in the same way as the visible O vacancies on $A_1$ and $B_1$ sites seen in (a).

Supplemental Material for

# Correlation of Fe-based Superconductivity and Electron-Phonon Coupling in an FeAs/Oxide Heterostructure

Seokhwan Choi[1], Steven Johnston[2], Won-Jun Jang[1,3†], Klaus Koepernik[4], Ken Nakatsukasa[2], Jong Mok Ok[5], Hyun-Jung Lee[5], Hyun Woo Choi[1], Alex Taekyung Lee[6], Alireza Akbari[5,7], Yannis K. Semertzidis[1,3], Yunkyu Bang[8], Jun Sung Kim[5], Jhinhwan Lee[1*]

*e-mail: jhinhwan@kaist.ac.kr

## 1. LDA + U calculations for $Sr_2VO_3FeAs$

We performed density functional theory calculations in the local density approximation [23] using the scalar relativistic mode of the Full Potential Local Orbital (FPLO) bandstructure code [24]. The density was converged using a 12×12×3 $k$-mesh for the tetrahedron integration method. The V 3d orbitals where treated within the LSDA+U approach, using the fully localized limit double counting scheme and a gross population number projector. We used $U = 6$ eV and $J = 1$ eV. We also checked that $U = 5$, $J = 1$ and $U = 6$, $J = 0.7$ do not alter the results in any essential way in the relevant energy region. We chose an arrangement of V spins such that each V moment is ordered ferromagnetically in plane and antiferromagnetically between planes. We did not investigate other magnetic patterns since there is already an extensive study in Ref. [25]. Our choice seems to be corroborated by experimental evidence [9,26]. In any case, the actual choice of magnetic order likely does not influence the results discussed here, provided that the V states are removed from the Fermi level. Given this magnetic pattern, two distinct solutions occur: one with a partially filled $d_{xy}$ orbital and one with an empty $d_{xy}$ orbital. The latter has occupied V $d_{yz}$ and $d_{xz}$ orbitals and a lower total energy and was therefore adapted for the data presented in this work. The calculated band structure is shown in Fig. S1(c). The Fermi surface for the undoped and doped cases is shown in Figs. S1(a) and S1(b), respectively. In this case, we simulated the doping by applying a rigid band shift.

We discuss electronic structure of $Sr_2VO_3FeAs$, which differs somewhat from the typical FeSCs, suggesting that the traditional nesting-based model may not hold in this material. *Ab initio* (LDA+U) calculations neglecting the onsite Hubbard interaction in the V atoms predict *both* Fe- and V-derived bands at the Fermi level $E_F$, while the Fe-derived bands have a topology similar to other FeSCs [27,28]. This, however, is not observed experimentally by ARPES, where no V-derived bands are found in proximity to $E_F$ and a shallow, rapidly dispersing electron-like band (the "$\alpha$" band) found at the Γ-point in Ref. [42] when using a certain polarization. Note that a similar band is hinted at in Fig. 2 of Ref. [41] but was not commented on. These observations can be accounted for in LDA+U calculations (Fig. S1(c)) [25], which remove the V-derived bands from the $E_F$ and predict a V local moment (consistent with bulk magnetization measurements [9,26]). This "$\alpha$" band is identified as the sharply dispersing band indicated by the arrow in Fig. S1(c). In this case, self-doping [9] and possibly additional self-energy effects beyond LDA+U, result in a shift of the bands relative to $E_F$ predicted by our LDA+U calculations. To mimic this effect, we have shifted the chemical potential upwards by 300 meV, resulting in the Fermi surface shown in Fig. S1(b). This band structure is qualitatively consistent with ARPES measurements [41,42] and our QPI measurements.

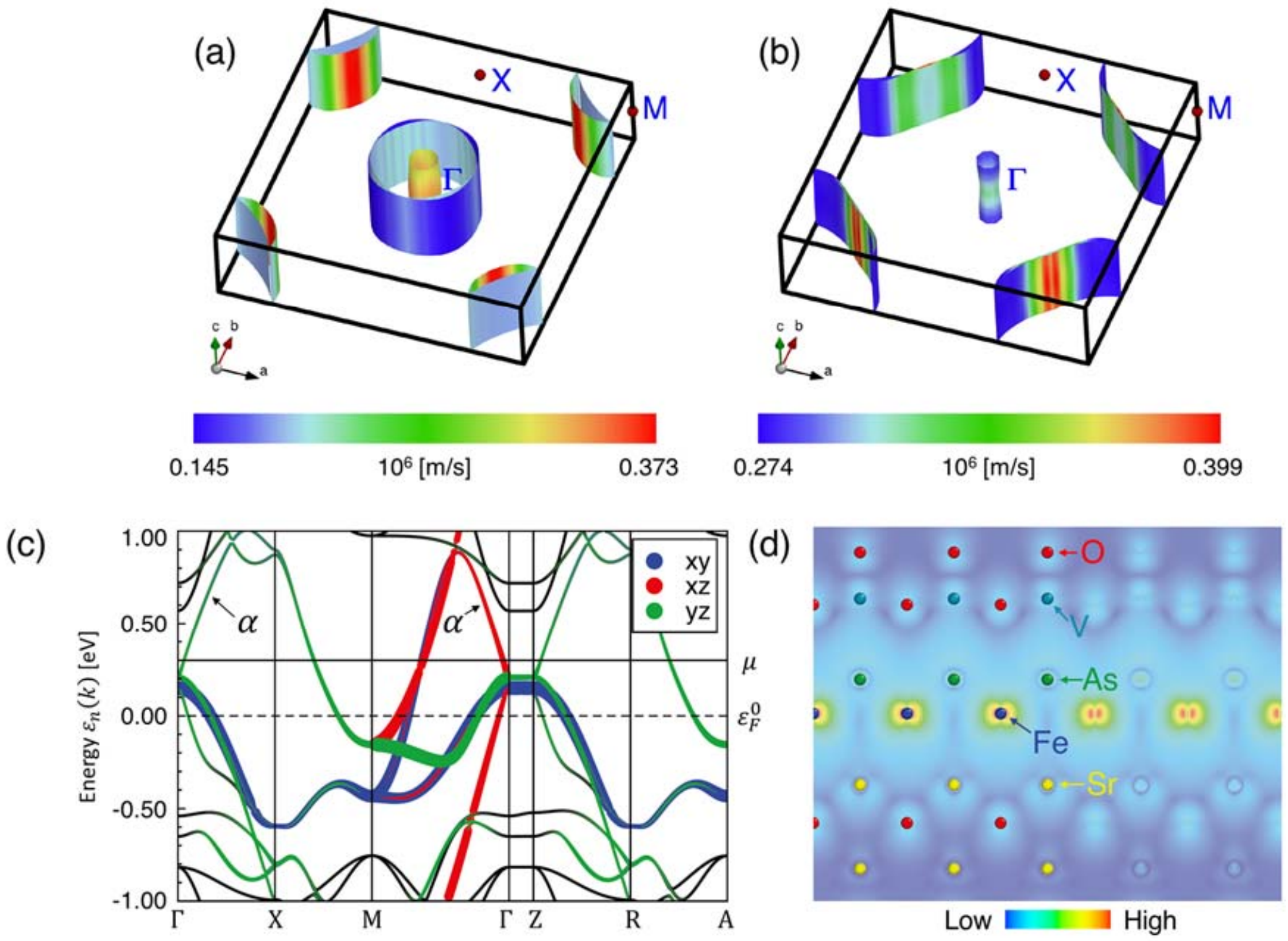

FIG. S1. The Fermi surfaces of $Sr_2VO_3FeAs$ computed within LDA + U, where the Hubbard U has been added to the V site. (a) shows the pnictide-like Fermi surface in the undoped system. (b) shows the Fermi surface in the system after self-doping, where the chemical potential has been raised by ~ 300 meV in order to match the observed Fermi surface topology. (c) Density functional theory (LDA+U) calculations for the band structure of $Sr_2VO_3FeAs$, where an onsite $U = 6$ eV was included for the V atoms. The thickness of the lines indicates the majority orbital content of the Fe-derive bands. The dashed line indicates the bare Fermi level $\varepsilon_F^0$ output by our calculations. Experiments indicate that self-doping effects and other many body corrections shift the chemical potential to the position indicated by the solid black line. (d) The spatial distribution of the electron density obtained by integrating the real space density from [-400, 400] meV. Most of the electron density is concentrated in the Fe-plane and the $(|\psi(\boldsymbol{r})|^2)^{\frac{1}{4}}$ is plotted to highlight density in the oxide layer.

## 2. Additional information on QPI signatures of superconductivity and e-ph coupling in $Sr_2VO_3FeAs$

In order to understand the nature of the various spectroscopic features found in the QPI measurement, we acquired QPI data with different temperature and magnetic field conditions as shown in Fig. S2 below. The narrow waists of the α-α band at the Fermi level are weakened at application of magnetic field of 7 T (Fig. S2(b)) and almost completely disappears at 40 K (Fig. S2(c)). The replica bands are persistent up to 60 K.

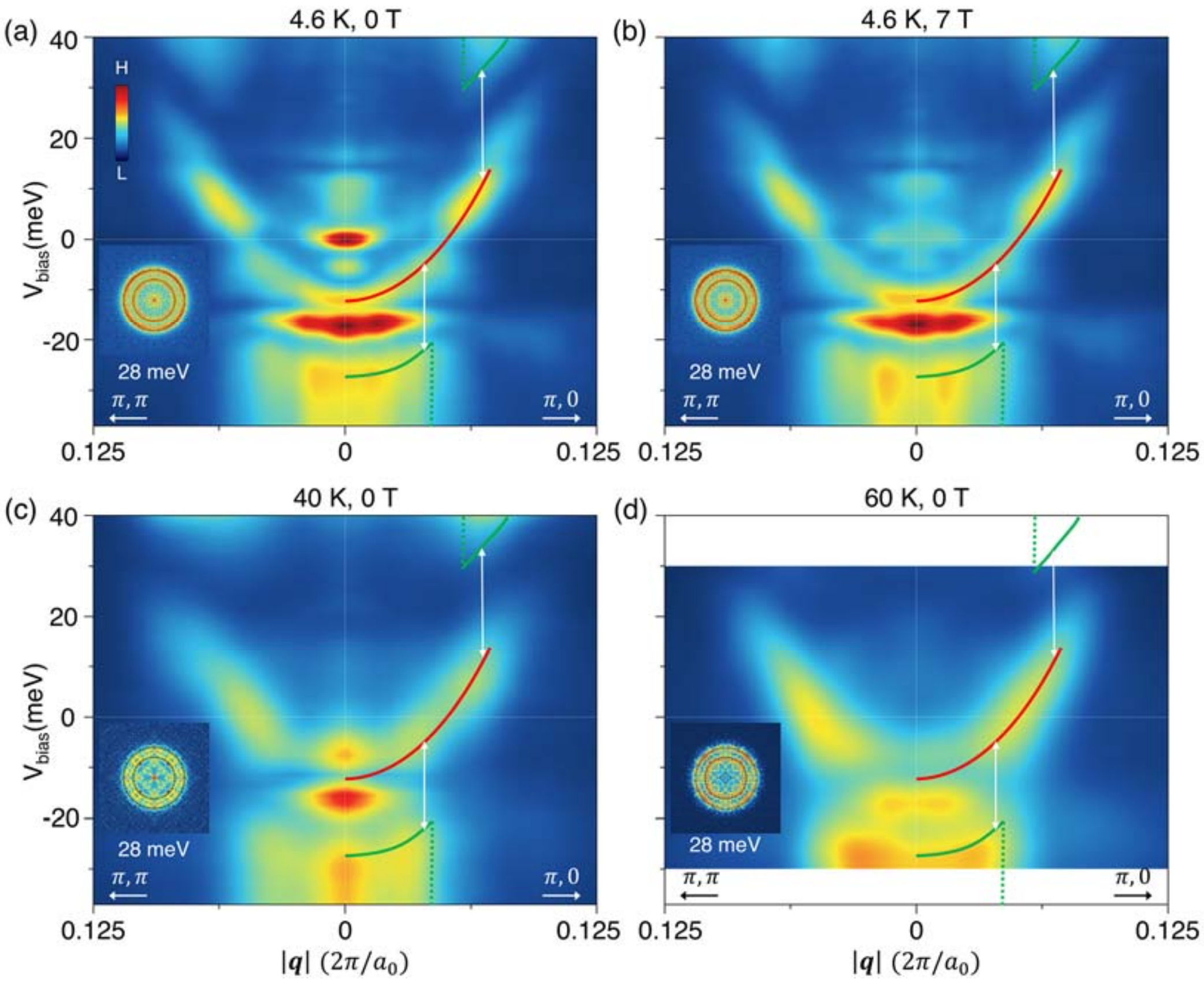

FIG. S2. The temperature and magnetic field dependence of QPI dispersions in $Sr_2VO_3FeAs$. The QPI cross-section images (a) at 4.6 K and 0 T, (b) at 4.6 K and 7 T, (c) at 40 K and 0 T and (d) at 60 K and 0 T. All the insets are constant energy (28 meV) QPI patterns. The insets are the representative empty state *q*-space images at +28 meV for each data, showing the replica and the original bands as two rings with different diameters. Figures S13 and S14 show the *q*-space images in other energy layers.

We can exclude the possibility that the replica bands could be due to tunneling into and out of virtual states (replica bands) coupled with real states (original bands) via emission of a forward-scattering phonon. This inelastic tunneling scenario, however, cannot explain the complex QPI band segment shifts and their correlations with superconducting gap, while they are naturally explained in terms of the dressing of the Fe electrons by the e-ph interaction as demonstrated here. Moreover, inelastic tunneling would not explain the gradual and complete cross-over of the original band and the replica band at a finite positive energy (~25 meV), which is reproduced only with the self-consistent calculation with high-order e-ph interactions [45].

## 3. Masked QPI analysis based on the superconducting gap ($\varDelta$) map

The spatial inhomogeneity of $\varDelta$ (superconducting gap) with characteristic length scale smaller than the QPI field of view allows for a $\varDelta$-controlled QPI analysis from a single QPI measurement, where all the other measurement parameters and defect distribution remains identical.

We measured the *dI/dV* map and extracted a simultaneous $\varDelta$ map by locating the coherence peak positions for each tunneling spectrum, as shown in Fig. S3(a). The field of view is divided into two regions with equal total areas based on the sign of $\varDelta - m(\varDelta)$ where $m(\varDelta)$ is the median value of the $\varDelta$ distribution. The $\varDelta^{+}$ ($\varDelta^{-}$) map in Figs. S3(b) and S3(c) contains mostly warm (cold) colors when plotted in the same color scale in Fig. S3(a). The rotationally averaged QPI dispersions obtained by Fourier transformation of $dI/dV$ images taken over only one of the two masked regions show slightly different sub-band positions and intensities, as shown in Fig. S5. This observation is consistent with positively correlated superconducting gap $(\varDelta)$ and electron-phonon coupling strength $\left(g_{ph}^{2}\right)$ as expected from the theory of band renormalization by the electron-phonon interaction (whose details are explained in Section 4).

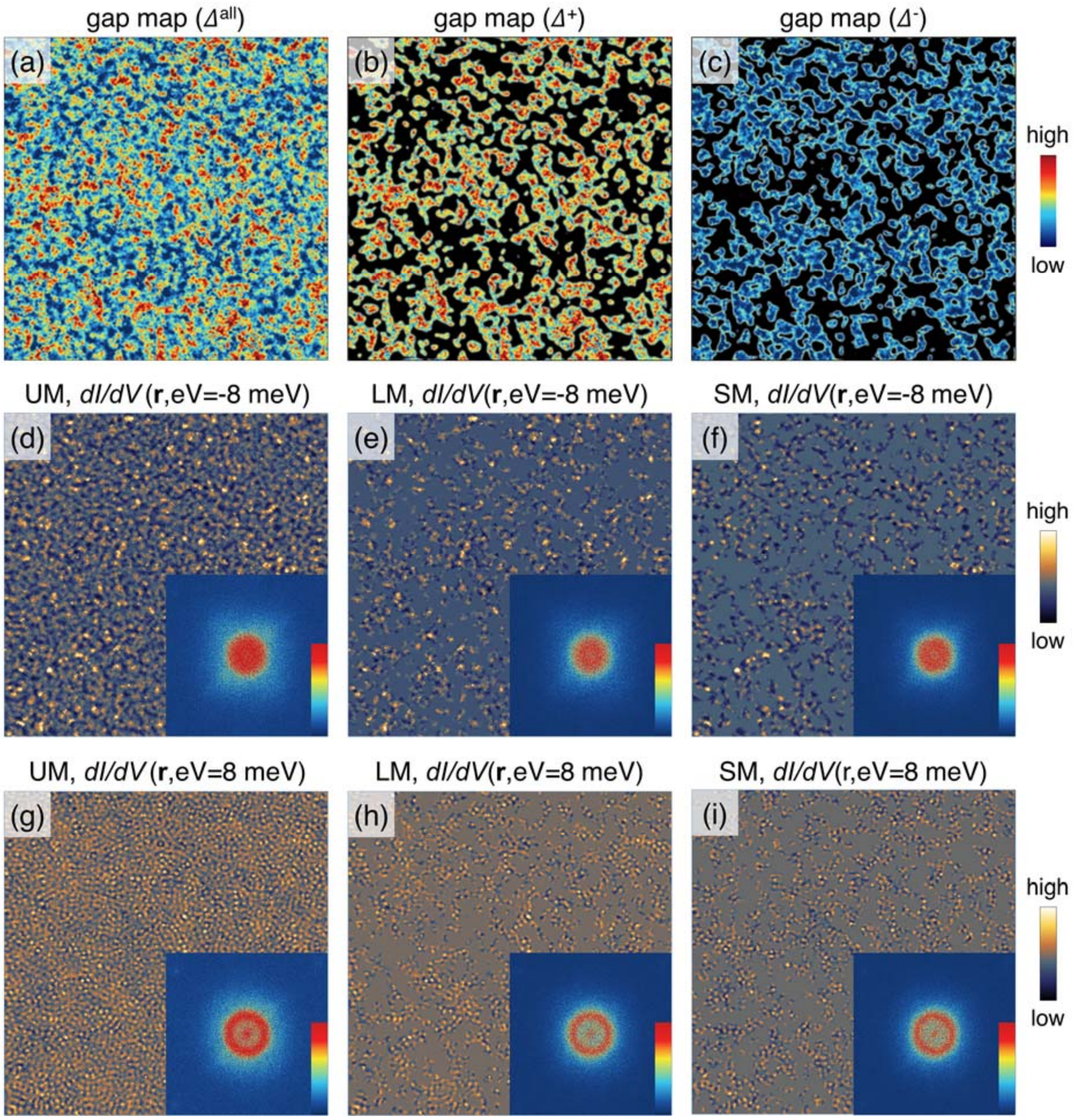

FIG. S3. (a) The superconducting gap ($\Delta$) map (300 nm × 300 nm) obtained simultaneously with the *dI/dV* map in Figs. 2 and 3. (b) The $\Delta^{+}$ mask with $\Delta > m(\Delta)$ (applied to the $\Delta$ map). (c) The $\Delta^{-}$ masks with $\Delta < m(\Delta)$ (applied to the $\Delta$ map). (d) Unmasked (UM) real space *dI/dV*(**r**, eV=-8 meV) map. (e) *dI/dV*(**r**, eV=-8 meV) in area $\Delta^{+}$(LM, Large gap Mask). (c) *dI/dV*(**r**, eV=-8 meV) in area $\Delta^{-}$(SM, Small gap Mask). (g)-(i) Images corresponding to (d)-(f) for +8 meV bias condition.

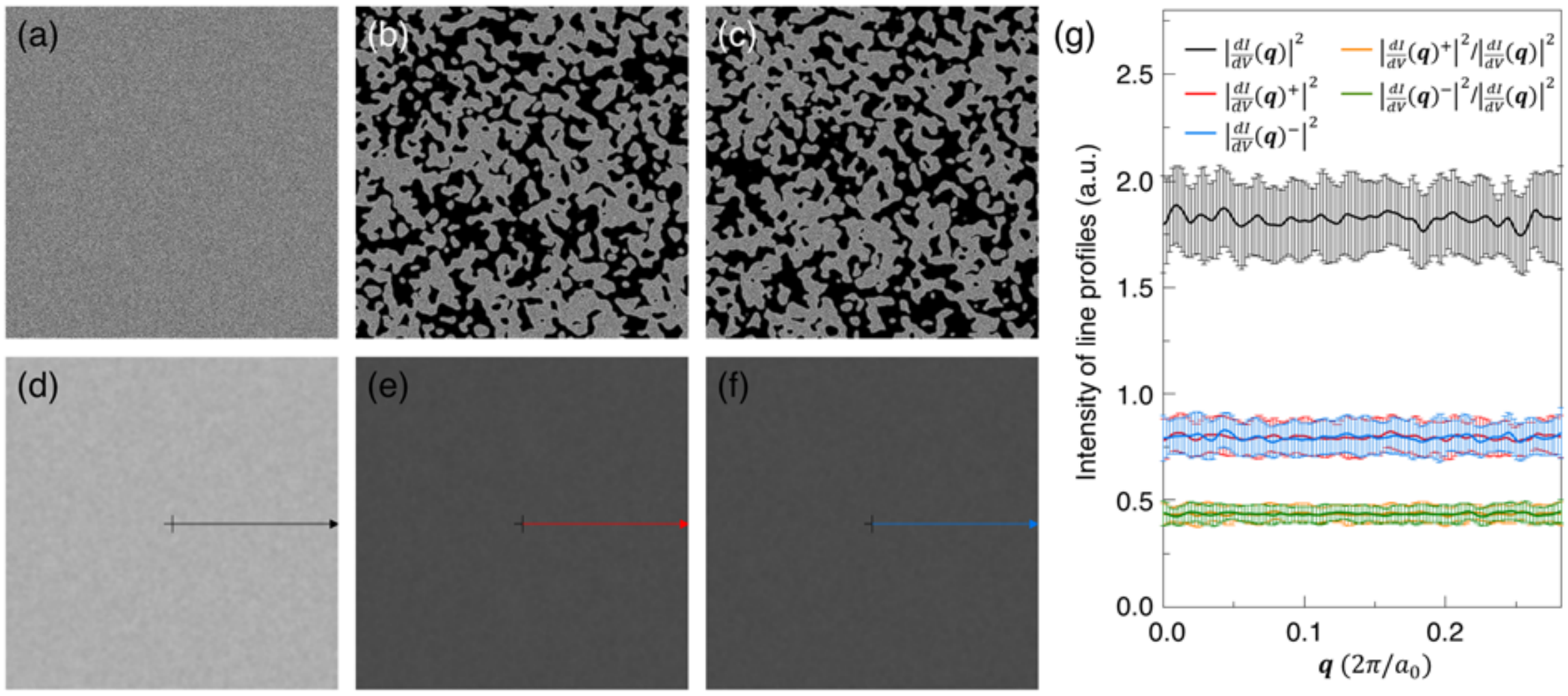

FIG. S4. (a) Virtual *dI/dV* image (512×512) with random numbers that contains nearly uniform ***q***-space intensity. (b),(c) $\Delta^+$ (b) and $\Delta^-$ (c) masks multiplied to (a). (d)-(f) Fourier-transformed images of (a)-(c) where the cross mark at the center corresponds to ***q***=0. (g) Line profiles along the three colored arrows in (d)-(f). It is noted that the ***q***-space responses introduced by application of the masks (the ratios of masked profiles divided by the unmasked profile) are virtually uniform for all possible ***q***.

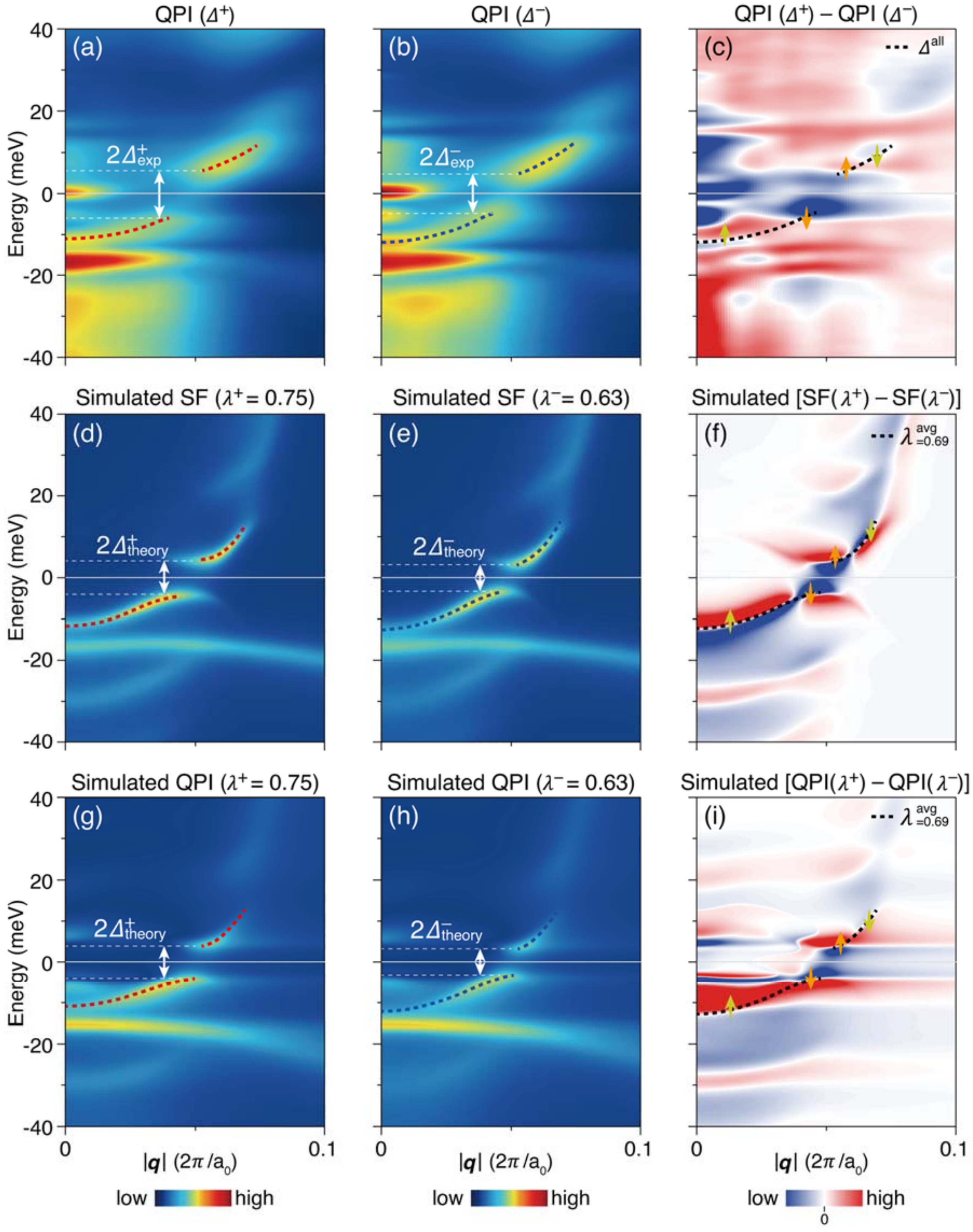

FIG. S5. (a),(b) Images of $dI/dV(|\boldsymbol{q}|, V)$ evaluated on two intertwined areas obtained by masking for gap value higher (a) and lower (b) than the median gap value. The red (blue) dashed curves indicate the trace of local maxima found in EDC curves of the data in (a)((b)). (c) The difference image of (a) and (b), where red (blue) color corresponds to the $dI/dV(|\boldsymbol{q}|, V)$ area with the value in (a) being larger (smaller) than the value in (b). The black dashed lines

indicate the EDC maxima of unmasked original $dI/dV(|\boldsymbol{q}|, V)$ data. (d)-(i) Images corresponding to (a)-(c) for the calculated spectral function (d)-(f) and QPI intensity (g)-(i) for parameters $\lambda^{-}$= 0.63 and $\lambda^{+}$= 0.75. From (c), (f) and (i), the band shifts of the filled and empty states α band segments toward the Fermi level (yellow arrows) are visible as the superconducting gap is increased (orange arrows).

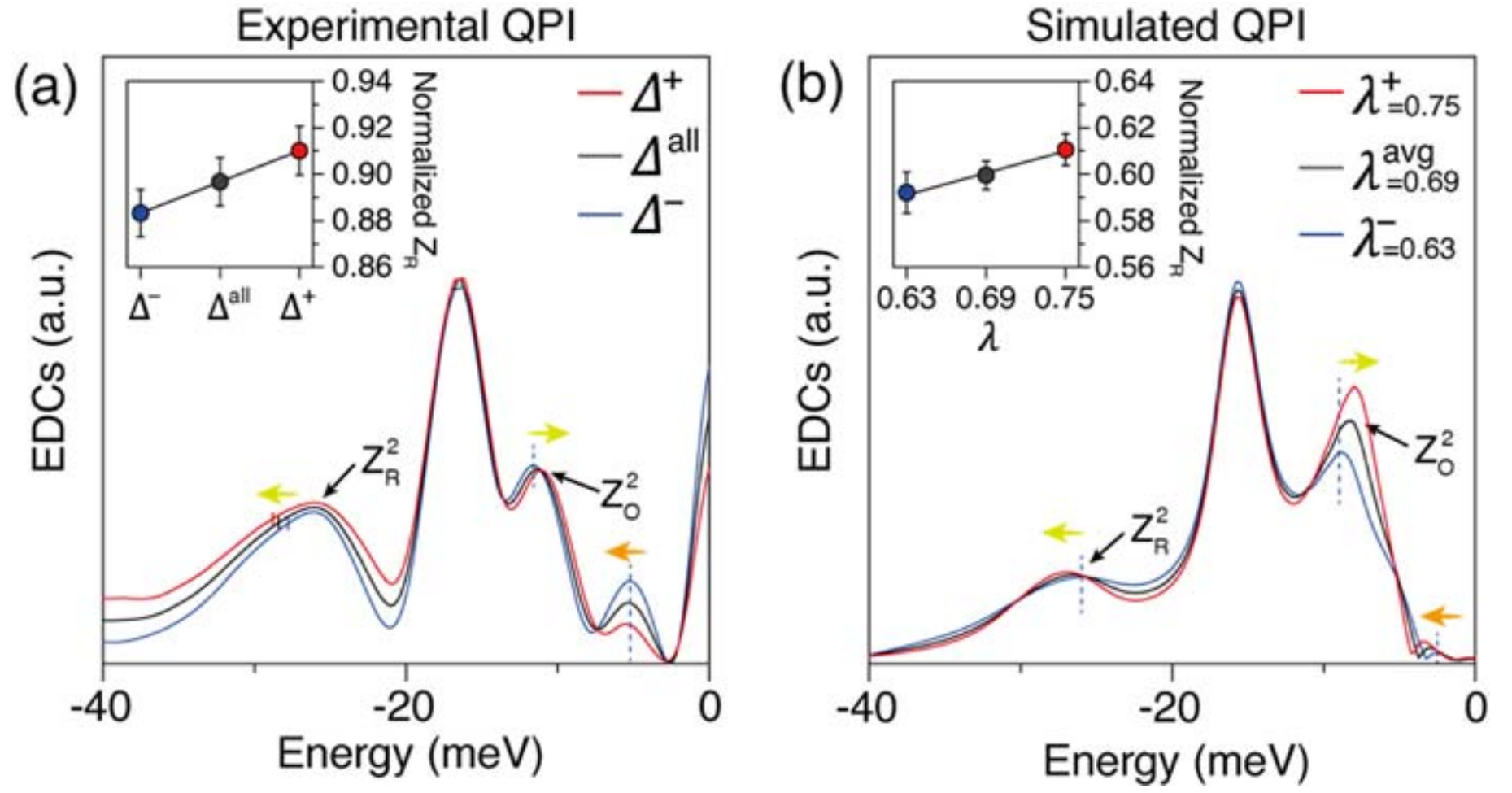

FIG. S6. (a) The EDC profiles of the masked QPI's shown in Fig. 3(a) ($\Delta^{+}$) and 3(b) ($\Delta^{-}$), and the unmasked QPI ($\Delta^{\mathrm{all}}$, equivalent to the sum of Fig. 3(a) and 3(b)), taken at $\boldsymbol{q}$ = 0.4$\boldsymbol{q_F}$. $Z_O$ and $Z_R$ are the spectral weights corresponding to the original ($\alpha$) and replica ($\alpha''$) bands, respectively. The inset shows the trend of normalized $Z_R$=$Z_R$/<$Z_O$> measured in the $\Delta^{-}$, $\Delta^{\mathrm{all}}$ and $\Delta^{+}$ regions at $\boldsymbol{q}$ = 0.4$\boldsymbol{q_F}$. (b) Plots corresponding to (a), but obtained from the simulated data of QPI shown in Figs. 3(c)-(d) and Figs. S5(g)-(i), calculated at three representative $\lambda$ values. The yellow arrows in (a) and (b) indicate that band shifts characteristic of larger $\Delta$ and $\lambda$.

## 4. Model for the electronic structure and QPI maps

We computed the single particle spectral function by self-consistently solving the momentum dependent Eliashberg equations, as outlined in Refs. [7,45]. The QPI intensity was computed using the T-matrix formalism as outlined in Ref. [45]. For our calculations, the range of the electron-phonon interaction was set by $q_0 = 0.1/a_0$. The phonon frequency was $\Omega = 10$ meV. The strength of the e-ph interaction was adjusted to give the $\lambda$ values quoted in the main text (which corresponds to $\lambda_m$ in the notation of Refs. [7,45]). The bare band structure was modeled with a single cosine-like band $\varepsilon(k) = -2t[\cos(k_x a_0) + \cos(k_y a_0)] - \mu_0$, with $t = 0.380$ eV and $\mu_0 = -1.485$ eV. The self-energy calculations were carried out self-consistently on a dense 256×256 momentum grid. The energy grid was set to a spacing of 0.25 meV. To produce the spectral function and QPI plots shown in the main text, the self-energy was interpolated onto a momentum grid four times denser using a cubic spline. The higher resolution spectral functions were then used to evaluate the QPI intensity, assuming a weak potential impurity scatterer with $V = 10$ meV.

## 5. Simplified picture: Spectral function formulation with self-energies within Migdal's first order approximation

The essential aspects of the Fermi-level-symmetric α band shifts due to increased electron-phonon coupling strength $g_{ph}^2$ and the increase in the replica band intensity can be explained as follows within the single-phonon approximation.

Using the electron-phonon coupling function $\hat{g}_{ph}^2(\boldsymbol{q}) = g_{ph}^2 N\delta(\boldsymbol{q})\hat{\tau}^0 \otimes \hat{\sigma}^3$, the lowest-order self-energy can be written as

$$\Sigma_{ph}^{(1)}(\boldsymbol{k}, i\omega_m) = g_{ph}^2 \left[\frac{\gamma^+(E_{\boldsymbol{k}})}{i\omega_m - E_{\boldsymbol{k}} - \Omega_{ph}} + \frac{\gamma^-(E_{\boldsymbol{k}})}{i\omega_m - E_{\boldsymbol{k}} + \Omega_{ph}}\right] \tag{S5.1}$$

where N is the number of momentum points, $g_{ph}^2$ characterizes the strength of the coupling, $\gamma^+(E_{\boldsymbol{k}}) = 1 - n_f(E_{\boldsymbol{k}}) + n_b(\Omega_k) \approx \Theta(E_{\boldsymbol{k}} - \mu)$ and $\gamma^-(E_{\boldsymbol{k}}) = n_f(E_{\boldsymbol{k}}) + n_b(\Omega_k) \approx \Theta(\mu - E_{\boldsymbol{k}})$ are both approximated with a unit step function $\Theta(x)$ for the bias energy scale $|E_{\boldsymbol{k}} - \mu|$ and the phonon mode energy scale $\Omega_{ph}$ larger than the temperature scale (~ 0.40 meV at 4.6 K).

The fermion Green's function given by

$$G^{(1)}(\boldsymbol{k}, i\omega_m) = \frac{1}{i\omega_m - E_{\boldsymbol{k}} - \Sigma_{ph}^{(1)}(\boldsymbol{k}, i\omega_m)} \tag{S5.2}$$

can then appear differently for the filled and the empty states. For $E_{\boldsymbol{k}} > \mu$ (empty states), $\gamma^-(E_{\boldsymbol{k}}) \approx 0$ and $\gamma^+(E_{\boldsymbol{k}}) \approx 1$ in the zero temperature limit and the Green's function is approximately written as

$$G_{empty}^{(1)}(\boldsymbol{k}, i\omega_m) \approx \frac{i\omega_m - E_{\boldsymbol{k}} - \Omega_{ph}}{(i\omega_m - E_{\boldsymbol{k}})(i\omega_m - E_{\boldsymbol{k}} - \Omega_{ph}) - g_{ph}^2} \tag{S5.3}$$

In the weak-coupling regime where $4g_{ph}^2 \ll \Omega_{ph}^2$ is satisfied, this Green's function can be further approximated by

$$G^{(1)}_{empty}(\boldsymbol{k}, i\omega_m) \approx \frac{1-g_{ph}^2/\Omega_{ph}^2}{i\omega_m-\left(\mu+\sqrt{\varepsilon_{\boldsymbol{k}}^2+\Delta^2}-g_{ph}^2/\Omega_{ph}\right)} + \frac{g_{ph}^2/\Omega_{ph}^2}{i\omega_m-\left(\mu+\sqrt{\varepsilon_{\boldsymbol{k}}^2+\Delta^2}+\Omega_{ph}+g_{ph}^2/\Omega_{ph}\right)} \quad \text{(S5.4)}$$

where we replaced $E_{\boldsymbol{k}} - \mu = \sqrt{\varepsilon_{\boldsymbol{k}}^2 + \Delta_{\boldsymbol{k}}^2} \approx \sqrt{\varepsilon_{\boldsymbol{k}}^2 + \Delta^2}$.

Similarly for $E_{\boldsymbol{k}} < \mu$ (filled states), $\gamma^-(E_{\boldsymbol{k}}) \approx 1$ and $\gamma^+(E_{\boldsymbol{k}}) \approx 0$ in the same temperature limit and the Green's function is approximately written as

$$G^{(1)}_{filled}(\boldsymbol{k}, i\omega_m) \approx \frac{i\omega_m - E_k + \Omega_{ph}}{(i\omega_m - E_k)(i\omega_m - E_k + \Omega_{ph}) - g_{ph}^2} \quad \text{(S5.5)}$$

and in the same weak-coupling limit

$$G^{(1)}_{filled}(\boldsymbol{k}, i\omega_m) \approx \frac{1-g_{ph}^2/\Omega_{ph}^2}{i\omega_m-\left(\mu-\sqrt{\varepsilon_{\boldsymbol{k}}^2+\Delta^2}+g_{ph}^2/\Omega_{ph}\right)} + \frac{g_{ph}^2/\Omega_{ph}^2}{i\omega_m-\left(\mu-\sqrt{\varepsilon_{\boldsymbol{k}}^2+\Delta^2}-\Omega_{ph}-g_{ph}^2/\Omega_{ph}\right)} \quad \text{(S5.6)}$$

where we replaced $E_{\boldsymbol{k}} - \mu = -\sqrt{\varepsilon_{\boldsymbol{k}}^2 + \Delta_{\boldsymbol{k}}^2} \approx -\sqrt{\varepsilon_{\boldsymbol{k}}^2 + \Delta^2}$.

As shown in Eqn's S5.4 and S5.6, the effects of variations of $\Delta$ and $g_{ph}$ on band segment shift directions are opposite to each other but symmetric with respect to the Femi level, which is consistent with the experimental observations. Furthermore, the replica bands have relative intensities of $\lambda \equiv g_{ph}^2/\Omega_{ph}^2$ compared with the original band in this first order approximation. In our QPI experiment, we observed $g_{ph}^2/\Omega_{ph}^2 \approx 0.7$ and the weak-coupling assumption appears marginally valid, indicating the possible necessity of higher order approximation or strong coupling theory in a more quantitative analysis.

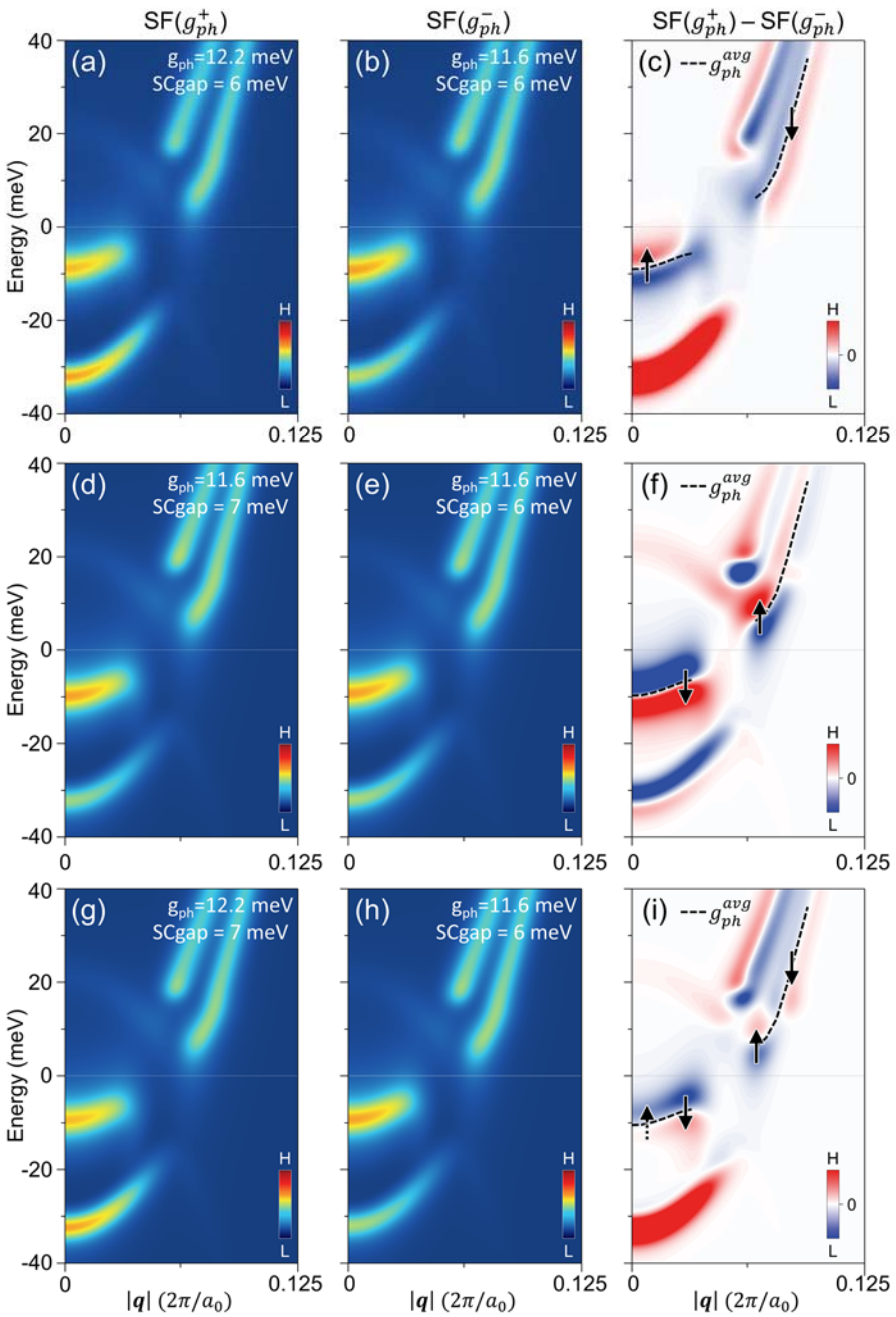

FIG. S7. Calculation results of shifts of spectral functions based on Eqs. (S5.4) and (S5.6) in first order e-ph interaction approximation. (a)-(c) For two different $g_{ph}$ values (11.6 meV and 12.2 meV) and fixed $\Delta$ (6 meV). (d)-(f) For fixed $g_{ph}$ value (11.6 meV) and two different $\Delta$ values (6 meV and 7 meV). (g)-(i) For two different $g_{ph}$ values (11.6 meV and 12.2 meV) and two covarying $\Delta$ values (6 meV and 7 meV).

## 6. LDA-based phonon mode calculations

Phonon calculations were performed using a supercell approach with the static finite displacement method. The real-space force constants of supercells were calculated within density-functional perturbation theory as implemented in the VASP code [29], using a 2×2×1 supercell. We used the PAW pseudopotential [30] and the revised version of the generalized gradient approximation (GGA) proposed by Perdew *et al.* (PBEsol) [31]. A plane wave basis with a kinetic energy cutoff of 500 eV and a k-point set generated by the 15×15×5 Γ-centered mesh were used. Phonon frequencies were calculated from the force constants using the PHONOPY code [32].

Crystal potential energy Φ is written as an analytic function of atomic displacements $u(lk)$ ($l$: labels of unit cells, $j$: labels of atoms in the unit cell),

$$\Phi = \Phi_0 + \sum_{lj} \sum_{\alpha} \Phi_{\alpha}(lj)\ u_{\alpha}(lj) + \frac{1}{2} \sum_{ll',jj'} \sum_{\alpha\beta} \Phi_{\alpha\beta}(lj, l'j')\ u_{\alpha}(lj)\ u_{\beta}(l'j') + \cdots,$$

where $\alpha$ and $\beta$ are Cartesian indices. $\Phi_0$, $\Phi_{\alpha}$, and $\Phi_{\alpha\beta}$ are the zeroth, first, and second order force constants, respectively.

Force and second order force constants are given by $F_{\alpha}(lj) = -\frac{\partial \Phi}{\partial u_{\alpha}(lj)}$ and $\frac{\partial^2 \Phi}{\partial u_{\alpha}(lj) \partial u_{\beta}(l'j')} = -\frac{\partial F_{\beta}(l'j')}{\partial u_{\alpha}(lj)}$, respectively. Within the harmonic approximation, the dynamical matrix is obtained by the following equation $D(\boldsymbol{q})\hat{\boldsymbol{\varepsilon}}_{\boldsymbol{q},\nu} = \omega_{\nu,\boldsymbol{q}}^2 \hat{\boldsymbol{\varepsilon}}_{\boldsymbol{q},\nu}$ with $D_{jj'}^{\alpha\beta}(\boldsymbol{q}) = \sum_{l'} \frac{\Phi_{\alpha\beta}(0j, l'j')}{\sqrt{m_j m_{j'}}} e^{i\boldsymbol{q}\cdot[r(l'j') - r(0j)]}$, where $m_j$ is the mass of the $j$-th atom, $\boldsymbol{q}$ is the wave vector, and $\nu$ is the band index.

## 7. Microscopic origin of electron-phonon coupling by interfacial phonon modes

The electron-phonon coupling can be regarded as primarily due to the modulation of the periodic potential felt by the itinerant FeAs electrons due to the oscillation of nearby ions. Depending on the detailed phase relationship of oscillation of various ions in the unit cell, some of the phonon modes couple more strongly to the itinerant electrons than others. Given the displacement eigenvector for each phonon mode, we can estimate the coupling strength of each mode as outlined in the following argument. We can assume that over the lateral dimension of each unit cell the itinerant Fe electrons are uniformly distributed in the Fe layer and the ions in the nearby $Sr_2VO_3$ layer can be modelled as point charges, which are assigned their nominal valence states ($q_{Sr}$=2$e$, $q_V$=3$e$, $q_O$= –2$e$). Taking into account of the structural symmetry of the ions in the unit cell of $Sr_2VO_3FeAs$, the Coulomb potential energy between the itinerant Fe electrons and all the ions is:

$$V_{ei} = \sum_{j}^{N_c}\sum_{l}^{N} V_{ei}\left(\boldsymbol{r}_e - \boldsymbol{R}_{l,j} - \boldsymbol{u}_{l,j,\boldsymbol{q},v}\right)$$

$$= -\sum_{j}^{N_c}\sum_{l}^{N}\frac{eq_j}{4\pi\varepsilon_{\parallel}}\frac{1}{\sqrt{\frac{\varepsilon_{\perp}}{\varepsilon_{\parallel}}\left(\left(X_{l,j}+x_{l,j,\boldsymbol{q},v}\right)^2+\left(Y_{l,j}+y_{l,j,\boldsymbol{q},v}\right)^2\right)+\left(Z_{l,j}+z_{l,j,\boldsymbol{q},v}\right)^2}} \quad \text{(S7.1)}$$

Here, $\varepsilon_{\parallel}$ and $\varepsilon_{\perp}$ are the dielectric constants parallel and perpendicular to the FeAs plane, $N$ is the total number of unit cells in the crystal, and $N_c$ is the number of ions in a unit cell. $\boldsymbol{u}_{l,j,\boldsymbol{q},v}$ is the displacement of the $j$-th ion in the perovskite layer in the $l$-th unit cell and $q_j$ is the point charge for the $j$-th ion. $\boldsymbol{R}_{l,j} = \boldsymbol{R}_l + \boldsymbol{R}_j$ is the position vector for the $j$-th ion in the $l$-th unit cell. The in-plane dielectric constant can be enhanced ($\varepsilon_{\parallel} \gg \varepsilon_{\perp}$) due to the screening by the nearly free electrons confined in the FeAs layer [4].

The pure oscillatory part of the potential energy is given by

$$\tilde{V}_{ei}(\boldsymbol{q},\nu) = V_{ei}\{\boldsymbol{u}_{\boldsymbol{q},\nu}\} - V_{ei}\{0\} \approx -\sum_j^{N_c}\sum_l^N \frac{eq_j}{4\pi\varepsilon_0}\frac{\kappa_\perp}{\kappa_\parallel}\frac{\frac{1}{\kappa_\parallel}(X_{l,j}x_{l,j,\boldsymbol{q},\nu}+Y_{l,j}y_{l,j,\boldsymbol{q},\nu})+\frac{1}{\kappa_\perp}(Z_{l,j}z_{l,j,\boldsymbol{q},\nu})}{\left|\frac{\kappa_\perp}{\kappa_\parallel}\left((X_{l,j})^2+(Y_{l,j})^2\right)+(Z_{l,j})^2\right|^{\frac{3}{2}}} \quad \text{(S7.2)}$$

Here $\kappa_\parallel = \varepsilon_\parallel/\varepsilon_0$ and $\kappa_\perp = \varepsilon_\perp/\varepsilon_0$ are the relative dielectric constants and $\boldsymbol{u}_{l,j,\boldsymbol{q},\nu} = (x_{l,j,\boldsymbol{q},\nu}, y_{l,j,\boldsymbol{q},\nu}, z_{l,j,\boldsymbol{q},\nu})$ is a complex vector containing the physical amplitude and the relative phase of each ion's motion for a given phonon mode $(\boldsymbol{q},\nu)$.

When the phonon energy is uniformly distributed over a crystal with $N$ unit cells, the expectation value of the total mechanical energy for each mode is quantized as

$$N\sum_j \langle\frac{1}{2}m_j\omega_{\boldsymbol{q},\nu}^2\left|\boldsymbol{r}_j(\boldsymbol{q},\nu)\right|^2\rangle + N\sum_j \langle\frac{1}{2}m_j\left|\boldsymbol{v}_j(\boldsymbol{q},\nu)\right|^2\rangle$$

$$= N\sum_j \langle m_j\omega_{\boldsymbol{q},\nu}^2\left|\boldsymbol{r}_j(\boldsymbol{q},\nu)\right|^2\rangle = \hbar\omega_{\boldsymbol{q},\nu} \quad \text{(S7.3)}$$

The physical displacement $\boldsymbol{r}_j(\boldsymbol{q},\nu) = \beta\boldsymbol{\varepsilon}_j(\boldsymbol{q},\nu)$ is constrained by the normalized displacement and the amplitude factor $\beta$ is determined by $N\sum_j \langle m_j\omega_{\boldsymbol{q},\nu}^2\beta^2\left|\boldsymbol{\varepsilon}_j(\boldsymbol{q},\nu)\right|^2\rangle = \hbar\omega_{\boldsymbol{q},\nu}$ , or $\beta^2 = \frac{\hbar}{N\omega_{\boldsymbol{q},\nu}\sum_j m_j\left|\boldsymbol{\varepsilon}_j(\boldsymbol{q},\nu)\right|^2} = \frac{\hbar}{N\omega_{\boldsymbol{q},\nu}\overline{m}_{\boldsymbol{q},\nu}}$ . Here $\overline{m}_{\boldsymbol{q},\nu} \equiv \sum_j m_j\left|\boldsymbol{\varepsilon}_j(\boldsymbol{q},\nu)\right|^2$ is the effective mass for each mode, close to the mass with the largest oscillation amplitude. Therefore the quantized amplitude is expressed in terms of the normalized displacement as

$$\boldsymbol{u}_{l,j,\boldsymbol{q},\nu} = \sqrt{\frac{\hbar}{N\omega_{\boldsymbol{q},\nu}\overline{m}_{\boldsymbol{q},\nu}}}\hat{\boldsymbol{\varepsilon}}_{j,\boldsymbol{q},\nu}e^{i\boldsymbol{q}\cdot\boldsymbol{R}_l} \quad \text{(S7.4)}$$

where complex normalized displacement denoted as $\hat{\boldsymbol{\varepsilon}}_{j,\boldsymbol{q},\nu}$, as supplied from the previous section, satisfies the normalization condition $\sum_j^{N_c}\left|\hat{\boldsymbol{\varepsilon}}_{j,\boldsymbol{q},\nu}\right|^2 = 1$.

From Eq. (S7.2) and (S7.3), the amplitude of the potential energy $\tilde{V}_{ei}(\nu,\boldsymbol{q})$ is written as

$\tilde{V}_{ei}(\nu,\boldsymbol{q}) \equiv g_{ph}(\nu,\boldsymbol{q}) =$

$$-\sum_j^{N_c}\sum_l^N \sqrt{\frac{\hbar}{N\omega_{\boldsymbol{q},\nu}\bar{m}_{\boldsymbol{q},\nu}}}\frac{eq_j}{4\pi\varepsilon_0}\frac{\kappa_\perp}{\kappa_\parallel}\frac{\frac{1}{\kappa_\parallel}(X_{l,j}x_{l,j,\boldsymbol{q},\nu}+Y_{l,j}y_{l,j,\boldsymbol{q},\nu})+\frac{1}{\kappa_\perp}(Z_{l,j}z_{l,j,\boldsymbol{q},\nu})}{\left|\frac{\kappa_\perp}{\kappa_\parallel}\left((X_{l,j})^2+(Y_{l,j})^2\right)+(Z_{l,j})^2\right|^{\frac{3}{2}}}e^{i\boldsymbol{q}\cdot\boldsymbol{R}_l} \quad \text{(S7.5)}$$

Therefore $g_{ph}(\nu,\boldsymbol{q})$ is a sharply peaked function at $\boldsymbol{q}=0$ (*i.e.,* zero momentum transfer) because it is basically the Fourier transform of $\frac{\frac{1}{\kappa_\parallel}(X_{l,j}x_{l,j,\boldsymbol{q},\nu}+Y_{l,j}y_{l,j,\boldsymbol{q},\nu})+\frac{1}{\kappa_\perp}(Z_{l,j}z_{l,j,\boldsymbol{q},\nu})}{\left|\frac{\kappa_\perp}{\kappa_\parallel}\left((X_{l,j}+x_{l,j,\boldsymbol{q},\nu})^2+(Y_{l,j}+y_{l,j,\boldsymbol{q},\nu})^2\right)+(Z_{l,j}+z_{l,j,\boldsymbol{q},\nu})^2\right|^{\frac{3}{2}}}$ which is a broadly peaked function of $\boldsymbol{R}_l$.

The total effective e-ph coupling strength $g_{ph}$ is

$g_{ph}^2 = \sum_{\boldsymbol{q}}\left|g_{ph}(\nu,\boldsymbol{q})\right|^2 =$

$$\frac{1}{N}\sum_{\boldsymbol{q}}\left|\sum_j^{N_c}\sum_l^N \sqrt{\frac{\hbar}{\omega_{\boldsymbol{q},\nu}\bar{m}_{\boldsymbol{q},\nu}}}\frac{eq_j}{4\pi\varepsilon_0}\frac{\kappa_\perp}{\kappa_\parallel}\frac{\frac{1}{\kappa_\parallel}(X_{l,j}x_{l,j,\boldsymbol{q},\nu}+Y_{l,j}y_{l,j,\boldsymbol{q},\nu})+\frac{1}{\kappa_\perp}(Z_{l,j}z_{l,j,\boldsymbol{q},\nu})}{\left|\frac{\kappa_\perp}{\kappa_\parallel}\left((X_{l,j})^2+(Y_{l,j})^2\right)+(Z_{l,j})^2\right|^{\frac{3}{2}}}e^{i\boldsymbol{q}\cdot\boldsymbol{R}_l}\right|^2 \quad \text{(S7.6)}$$

where the outer $\boldsymbol{q}$-sum is over the $N$ uniformly spaced points in the first Brillouin zone and exactly cancels out the $1/\sqrt{N}$-dependence of the quantized phonon amplitude in the thermodynamic limit.

In case $\omega_{\boldsymbol{q},\nu}$ and $\hat{\boldsymbol{\varepsilon}}_{j,\boldsymbol{q},\nu}$ have only small $\boldsymbol{q}$-dependences and are nearly uniform inside the sharp peak of $g_{ph}(\nu,\boldsymbol{q})$ centered at $\boldsymbol{q}=0$, and if the vibration of all ions are confined in z direction, we can further simplify

$$g_{ph}^2 = \sum_{\boldsymbol{q}}\left|g_{ph}(\nu,\boldsymbol{q})\right|^2 \sim \frac{1}{N}\sum_{\boldsymbol{q}}\left|\sum_j^{N_c}\sum_l^N \sqrt{\frac{\hbar}{\omega_{\boldsymbol{0},\nu}\bar{m}_{\boldsymbol{0},\nu}}}\frac{eq_j}{4\pi\varepsilon_\parallel}\frac{Z_{l,j}z_{l,j,\boldsymbol{0},\nu}}{\left|\frac{\varepsilon_\perp}{\varepsilon_\parallel}\left((X_{l,j})^2+(Y_{l,j})^2\right)+(Z_{l,j})^2\right|^{\frac{3}{2}}}e^{i\boldsymbol{q}\cdot\boldsymbol{R}_l}\right|^2 \quad \text{(S7.7)}$$

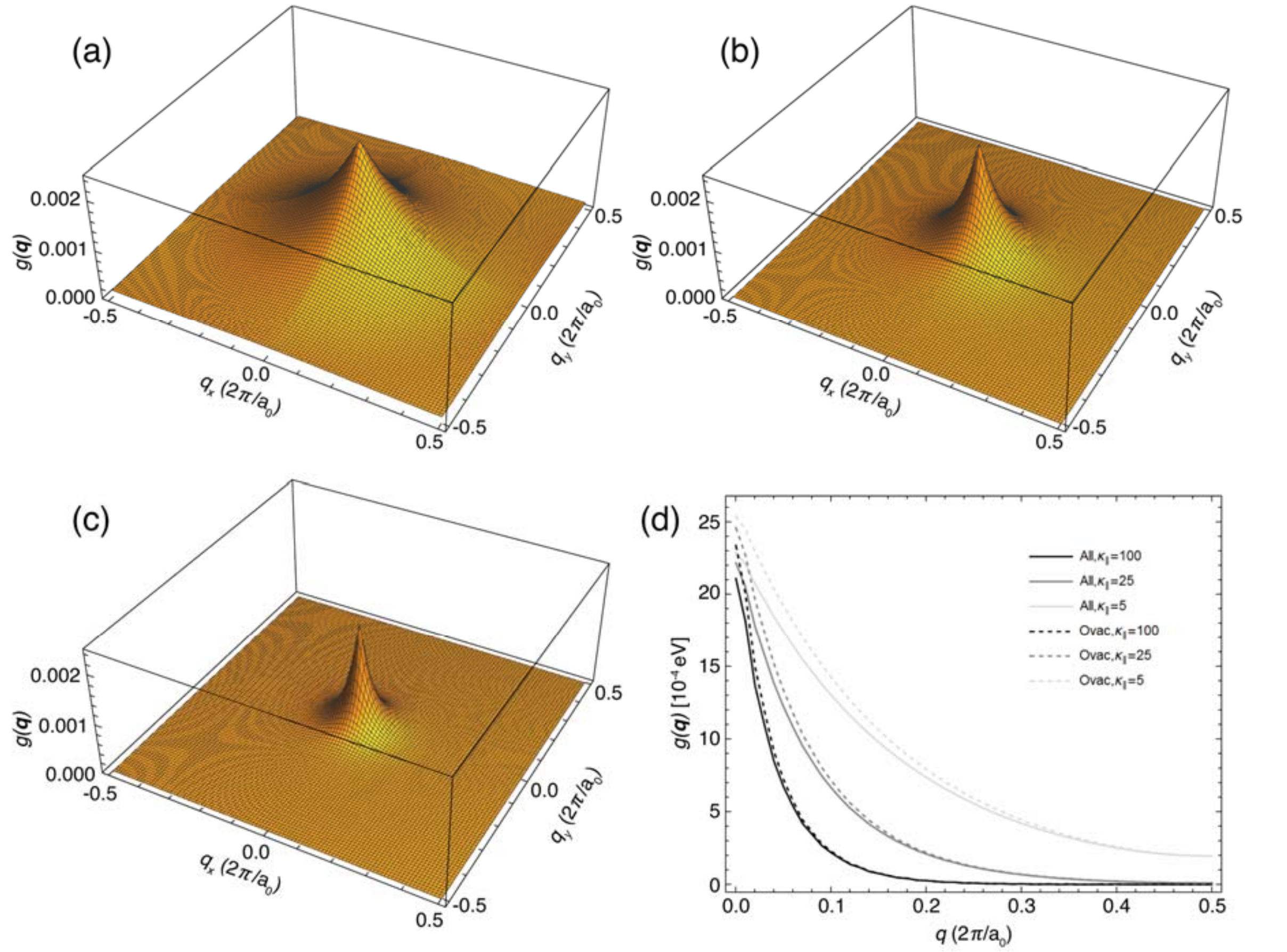

FIG. S8. The plots of $g_{ph}(\boldsymbol{q})$ evaluated for the symmetrical breathing mode of $Sr_2VO_3FeAs$. (a)-(c) $g_{ph}(\boldsymbol{q})$ plotted over the first BZ for $\kappa_\perp$ = 5 and (a) $\kappa_\parallel$ = 5, (b) $\kappa_\parallel$ = 25 and (c) $\kappa_\parallel$ = 100. The 2D grid indicates the sampling points with $N$ = $10^4$. (d) $g_{ph}(\boldsymbol{q})$ plotted over ΓX for the three $\kappa_\parallel$ values of (a)-(c). The solid curves correspond to the full $Sr_2VO_3FeAs$ structure and the dashed curves correspond to the increased e-ph coupling due to an oxygen ion in one of the $VO_2$ layers missing, modelled by setting the charge $q_j$=0 for the corresponding oxygen ion.

In order to estimate the coupling strength $g_{ph}(\boldsymbol{q})$ in case of $Sr_2VO_3FeAs$, the basic phonon mode parameters such as $\boldsymbol{\varepsilon}_j(\boldsymbol{q},\nu)$, $\omega(\boldsymbol{q},\nu)$ and $\boldsymbol{R}_j$ are extracted from first-principles calculations using Phonopy (Fig. S9). Currently it is difficult to estimate the exact values of the relative dielectric constants $\kappa_\perp$ and $\kappa_\parallel$. However, it is expected that $\kappa_\perp$ is close to the value for bulk perovskite while $\kappa_{\parallel}$ is much higher due to the screening by the itinerant electrons in the FeAs plane. Using $\kappa_{\perp} \approx 4.2$ [33] and $\kappa_{\parallel} \approx 140$ and the symmetrical breathing optical phonon mode at $\omega_{\boldsymbol{q}} \approx \Omega_{ph} \approx 13.1$ meV, the effective coupling strength $g_{ph} \approx 11$ meV results which is a fair approximation that makes the simulation result close to the experiment. We can then obtain an approximate estimate for the dimensionless coupling constant $\lambda$ using the approximation used elsewhere [7,19]

$$\lambda = \langle Re \frac{\partial \Sigma^{(1)}(k_F,\omega)}{\partial \omega}\Big|_{\omega=0} \rangle = \frac{g_{ph}^2}{\Omega_{ph}^2} \approx 0.7 \qquad \text{(S7.7)}$$

This same limit yields an estimate for the critical temperature $T_c$ [7],

$$T_c \approx \frac{\lambda}{2+3\lambda}\Omega_{ph} \approx 2.2 \text{ (meV)} \approx 26 \text{ (K)} \qquad \text{(S7.8)}$$

and a zero-temperature gap magnitude of

$$\Delta = 2k_B T_c \approx 4.4 \text{ (meV).} \qquad \text{(S7.9)}$$

Note that these estimates are based on the results derived in the limit of "perfect forward scattering" and serve as an analytical solution to guide our thinking. The gap values due to the interfacial phonon coupling quoted in the main text were obtained from the full solution to the momentum dependent Eliashberg equations, assuming an e-ph coupling with a finite width as outlined in Section S4. The simulation produces a smaller estimate for $T_c$ and $\Delta$ since it accounts for the full width of the interaction in momentum space (i.e. a finite $q_0$), which lowers the values relative to the limit of perfect forward scattering [7].

The significant increase by of the e-ph coupling near the measured O vacancy defects is due to the decrease of number of cations (O ions) partially cancelling the effect of most dominant Coulomb potential oscillation effect of the metallic anions (Sr and V ions). The average gap energy increase of 0.2 meV seen in Fig. 4(c) may be compared with the theoretically calculated upper limit of the gap energy increase of 0.33 meV with the assumption that the dimensionless electron-phonon coupling constant ($\lambda$) may be increased by 17 % near each O vacancy in the $VO_2$ layer, based on the frozen phonon calculation used for Fig. S8(d) and Eqs. S7.7-S7.9. The discrepancy could be reduced by application of a more realistic model of a single isolated defect embedded in a large calculation unit cell suitable for virtually isolated single defects as are the cases for the O vacancies found here.

Considering that the O vacancies on the other side of the top FeAs layer (on $A_2$ and $B_2$ sites shown in Fig. 4(e)) are not directly visible in our tunneling measurements, and from the significant increase in the average superconducting gap correlated with the visible O vacancies (on the same side of the top FeAs layer as the tip, i.e. on $A_1$ and $B_1$ sites), it is reasonable to be able to find some regions of large gap without visible O vacancies since they may correspond to the regions with many invisible O vacancies beneath the top FeAs layer.

At the same time, since e-ph interaction is only an additional pairing mechanism in this material, the gap inhomogeneity that determines the width of the gap histogram distribution may mostly come from other channels related to the spin or orbital based pairing which are not the topic of this paper.

TABLE SI. Lattice parameters and phonon mode parameters for the symmetrical breathing optical phonon mode ($\hbar\omega\left(\boldsymbol{q}=\boldsymbol{0},\nu\right)=$ 13.1 meV).

| | Equilibrium ionic positions ($\boldsymbol{R}_j$) [m] | | | Normalized ionic displacement for the phonon mode ($\boldsymbol{\varepsilon}_j(\boldsymbol{0},\nu)$) | | | | | | | Ionic mass | Ionic charge | |
|---|---|---|---|---|---|---|---|---|---|---|---|---|---|
| Ion ($j$) | X | Y | Z | Re(x) | Re(y) | Re(z) | Im(x) | Im(y) | Im(z) | $\left\lvert\boldsymbol{\varepsilon}_j(\boldsymbol{0},\nu)\right\rvert^2$ | $m_j$[amu] | $q_j$ | Lattice constant (nm) |
| **Sr1** | 1.34E-15 | 1.31E-15 | -2.84E-10 | -2.19E-03 | -2.19E-03 | -5.12E-01 | 0 | 0 | 5.90E-02 | **2.66E-01** | 87.6 | 2 $e$ | a |
| Sr2 | 1.50E-15 | 1.53E-15 | -6.49E-10 | 2.34E-03 | 2.34E-03 | -2.25E-01 | -3.00E-05 | -3.00E-05 | 3.45E-02 | 5.20E-02 | 87.6 | 2 $e$ | 0.39296 |
| **Sr3** | 1.96E-10 | 1.96E-10 | 2.84E-10 | 4.35E-03 | 4.35E-03 | 5.18E-01 | -7.25E-04 | -7.25E-04 | -3.33E-02 | **2.69E-01** | 87.6 | 2 $e$ | b |
| Sr4 | 1.96E-10 | 1.96E-10 | 6.49E-10 | -6.29E-03 | -6.29E-03 | 2.30E-01 | 1.07E-03 | 1.07E-03 | -6.68E-03 | 5.30E-02 | 87.6 | 2 $e$ | 0.39296 |
| V1 | 0 | 0 | 4.70E-10 | 1.59E-03 | 1.59E-03 | 3.17E-01 | -2.95E-04 | -2.95E-04 | -1.51E-02 | 1.01E-01 | 50.9 | 3 $e$ | c |
| V2 | 1.96E-10 | 1.96E-10 | -4.70E-10 | -4.74E-04 | -4.74E-04 | -3.10E-01 | -8.33E-05 | -8.33E-05 | 4.18E-02 | 9.76E-02 | 50.9 | 3 $e$ | 1.567319 |
| O1 | 3.11E-15 | 3.10E-15 | 6.69E-10 | 1.70E-03 | 1.70E-03 | 1.17E-01 | -3.01E-04 | -3.01E-04 | -6.73E-03 | 1.39E-02 | 16.0 | -2 $e$ | |
| O2 | 1.96E-10 | 1.96E-10 | -6.69E-10 | -2.02E-04 | -2.02E-04 | -1.15E-01 | -1.17E-04 | -1.17E-04 | 1.44E-02 | 1.33E-02 | 16.0 | -2 $e$ | |
| **O3** | 1.96E-10 | 7.81E-16 | 4.59E-10 | 1.11E-03 | 1.25E-03 | 1.71E-01 | -1.97E-04 | -2.07E-04 | -9.98E-03 | **2.92E-02** | 16.0 | -2 $e$ | |
| **O4** | 7.69E-16 | 1.96E-10 | 4.59E-10 | 1.25E-03 | 1.11E-03 | 1.71E-01 | -2.07E-04 | -1.97E-04 | -9.98E-03 | **2.92E-02** | 16.0 | -2 $e$ | |
| **O5** | 3.46E-16 | 1.96E-10 | -4.59E-10 | -3.21E-04 | -3.84E-04 | -1.68E-01 | -5.08E-05 | -3.54E-05 | 2.06E-02 | **2.85E-02** | 16.0 | -2 $e$ | |
| **O6** | 1.96E-10 | 3.46E-16 | -4.59E-10 | -3.84E-04 | -3.21E-04 | -1.68E-01 | -3.54E-05 | -5.08E-05 | 2.06E-02 | **2.85E-02** | 16.0 | -2 $e$ | |
| Fe1 | 1.96E-10 | 4.77E-16 | 0.00E+00 | -4.62E-04 | -1.16E-03 | -3.74E-03 | 1.46E-04 | 2.25E-04 | -1.65E-02 | 2.89E-04 | 55.8 | | |
| Fe2 | 5.13E-16 | 1.96E-10 | 0.00E+00 | -1.16E-03 | -4.62E-04 | -3.74E-03 | 2.25E-04 | 1.46E-04 | -1.65E-02 | 2.89E-04 | 55.8 | | |
| As1 | 7.57E-16 | 7.69E-16 | 1.26E-10 | 1.11E-04 | 1.11E-04 | 8.70E-02 | 7.24E-05 | 7.24E-05 | -3.14E-02 | 8.55E-03 | 74.9 | | |
| As2 | 1.96E-10 | 1.96E-10 | -1.26E-10 | -1.75E-03 | -1.75E-03 | -9.70E-02 | 2.54E-04 | 2.54E-04 | -1.51E-02 | 9.64E-03 | 74.9 | | |
| | | | | | | | | | | $\sum_j\left\lvert\boldsymbol{\varepsilon}_j(\boldsymbol{0},\nu)\right\rvert^2=1$ | $\sum_j m_j\left\lvert\boldsymbol{\varepsilon}_j(\boldsymbol{0},\nu)\right\rvert^2\equiv\bar{m}=69.9$ | | |

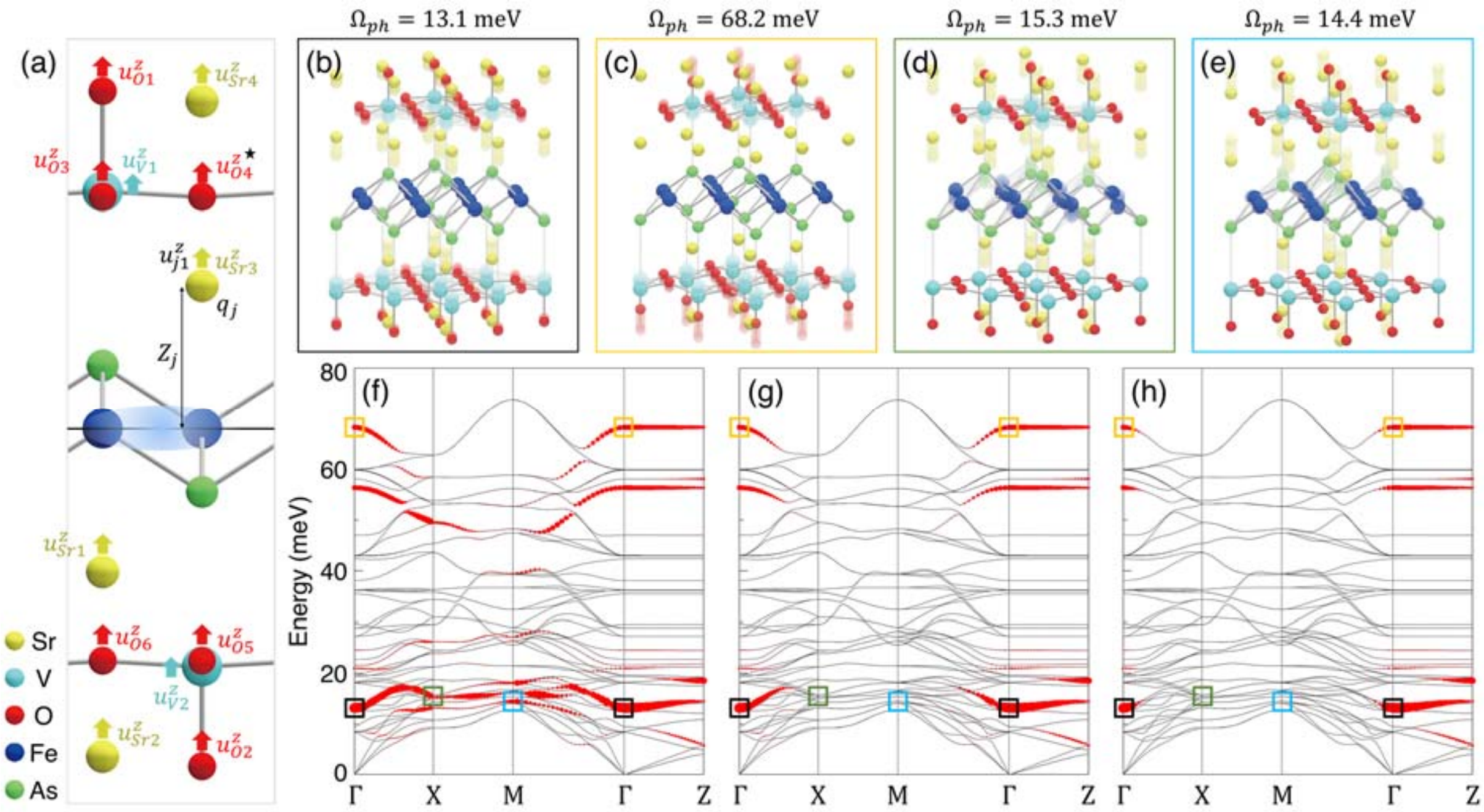

FIG. S9. (a) The Coulomb electron-ion interaction model for $Sr_2VO_3FeAs$ used to estimate the effective electron-phonon coupling strength for the itinerant electrons for a given ionic displacement eigenvector $\boldsymbol{\varepsilon}_j(\boldsymbol{q}, \nu)$ extracted from a phonon mode calculation for the $\nu$-th phonon mode. (b),(c) The two Γ phonon modes with the strongest Coulomb couplings: (b) the symmetrical breathing mode of the whole $Sr_2VO_3$ layers with respect to the FeAs layer and (c) the symmetrical "V to apical O dipole"-based optical phonon mode more confined to each $VO_2$ layer. (d),(e) The two BZ boundary modes with vanishing Coulomb coupling with itinerant Fe electrons due to cancellation of nearest neighbor contributions. (f)-(h) The phonon mode plots of $Sr_2VO_3FeAs$ with the red disc size indicating the Coulomb coupling, with three different momentum cut offs due to various $\frac{\kappa_\perp}{\kappa_\parallel}$ values. The energy and momentum of each of the phonon modes shown in (b) to (e) are marked on the phonon mode plots (f) to (h) inside boxes with corresponding colors.

## 8. Determination of the coherence peak position in presence of a large quadratic background

In case the coherence peaks coexist with large background tunneling spectrum, determining its true energy location is challenging. In most cases the coherence peaks can be modelled as a symmetric peak $g_0(V) = de^{-\left(\frac{V}{u}\right)^2}$ (peak center is set to zero without loss of generality) and the background can be modelled as a quadratic function of energy so that the tunneling spectrum is given by $g(V) \equiv dI/dV(V) = aV^2 + bV + c + g_0(V)$. Then we can determine the hidden peak position by locating the negative peak in its second derivative $g''(V) \equiv d^3I/dV^3(V)$, since $g''(V) = 2a + g_0''(V) = 2a - 2de^{-\left(\frac{V}{u}\right)^2}\frac{u^2-V^2}{u^4}$ is an even function with a strong negative peak located at $V = 0$, i.e. the center of the original hidden peak $g_0(V)$.

In practice, when the gap map pixel size is smaller than the coherence length, we can average each spectrum with those of surrounding 8 pixels to enhance the S/N before taking the polynomial fitting. We then can take the second derivative of the polynomial algebraically to determine its negative peak position.

Figure S10 shows some extreme cases where the local maxima of $g(V) \equiv dI/dV(V)$ (shown in blue) are located away (a-c) from those of $g_0(V)$ (shown in red) or even cannot be determined (d) due to the large background slopes. However, the local minima of $g''(V)$ (shown in green) are always located at the center position of $g_0(V)$.

Figure S11 shows how the algorithm is applied to a set of tunneling spectra taken along a line on $Sr_2VO_3FeAs$.

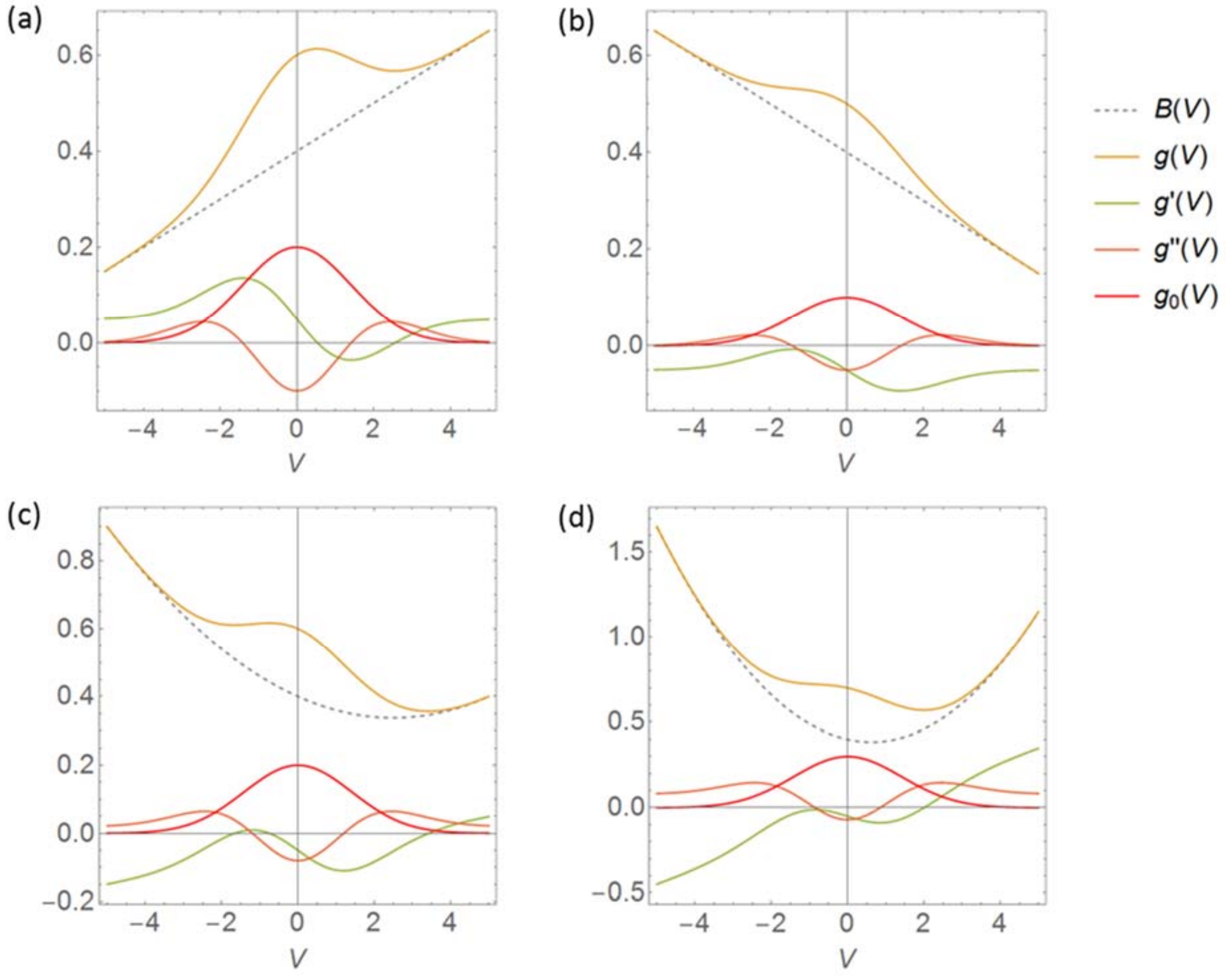

FIG. S10. Examples of hidden peak position extraction in case of strong quadratic background. The original tunneling spectrum $g(V) \equiv dI/dV(V)$ is modelled by $g(V) = aV^2 + bV + c + g_0(V)$ where $g_0(V) = de^{-\left(\frac{V}{u}\right)^2}$. The second derivative of $g(V)$ (i.e. $g''(V)$ shown in green) always results in a single negative peak at the position of the original hidden peak (i.e. $g_0(V)$ shown in red) no matter what the values of $a$, $b$ and $c$ are for the given $d$ and $u$. (a) {a,b,c,d,u}={0,0.05,0.4,0.2,2} (b) {0,-0.05,0.4,0.1,2} (c) {0.01,-0.05,0.4,0.2,2} (d) {0.04,-0.05,0.4,0.3,2}.

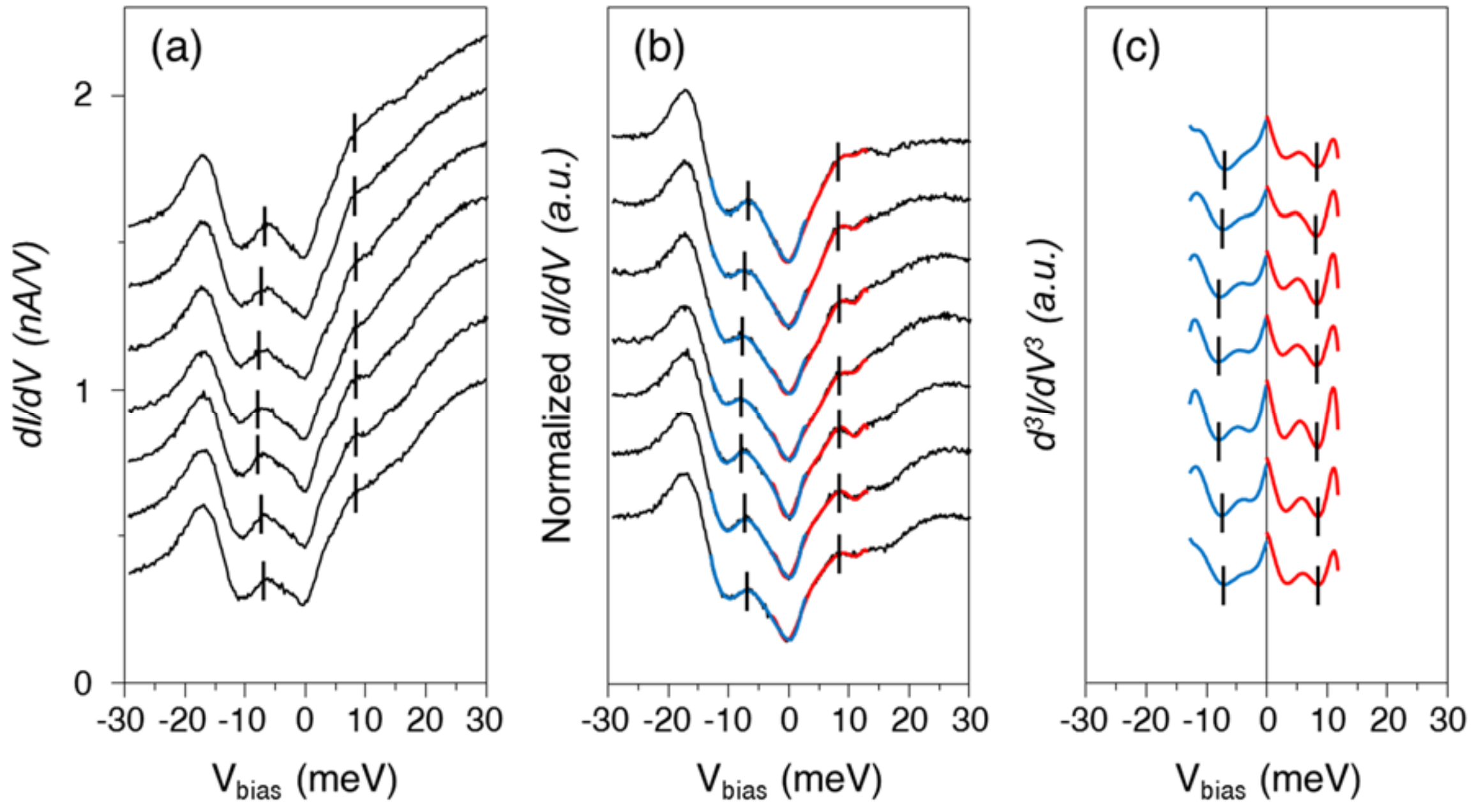

FIG. S11. Demonstration of the algorithm used to extract the coherence peak positions from actual tunneling spectra. (a) Raw *dI/dV* data taken over a line with 3-nm range. (b) *dI/dV* data with background subtraction applied using the 140 K tunneling spectrum. The short blue and red overlaid curves are 9$^{th}$ order polynomial fitted to the data in the negative ([-13,+3] meV) and positive ([-3,+13] meV) energy coherence peak regions separately. (c) Analytic 2$^{nd}$ order differentiation applied on the polynomials in (b), resulting in $d^3I/dV^3$ whose local minima correspond to the positions of the coherence peaks in the presence of strong background spectra as in this case. The short black vertical markers in all panels show the positions of the local minima found in (c).

## 9. Producing the *q*-vs-*E* cross-sectional images from the raw QPI map

Starting from the raw 512×512×201 *dI/dV* data set which are shown in Fig. S12 below at 12 representative energies, we have taken the Fourier transform of every energy layer and then applied 2-3 pixel radius Gaussian averaging in the *q* space before taking the cross-sections in two major high-symmetry axis directions. The Fig. S13 show raw Fourier-transformed images and the cross-sections in the main figures and in Fig. S14 are results of 2-pixel Gaussian averaging in *q* space.

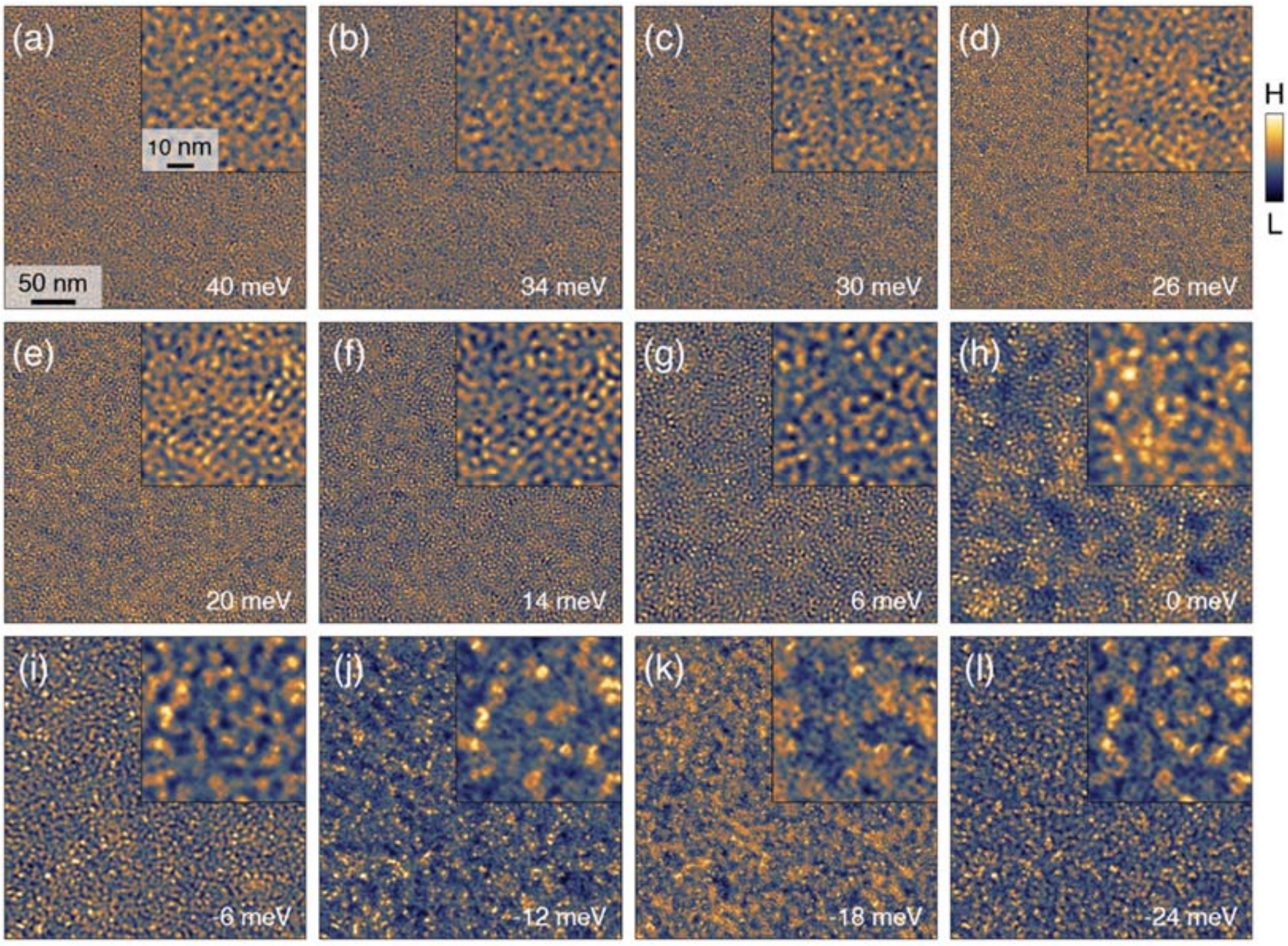

FIG. S12. Raw *dI/dV* data in real space. The inset in each panel is zoomed-in views of the corresponding large area real space data shown for clarity.

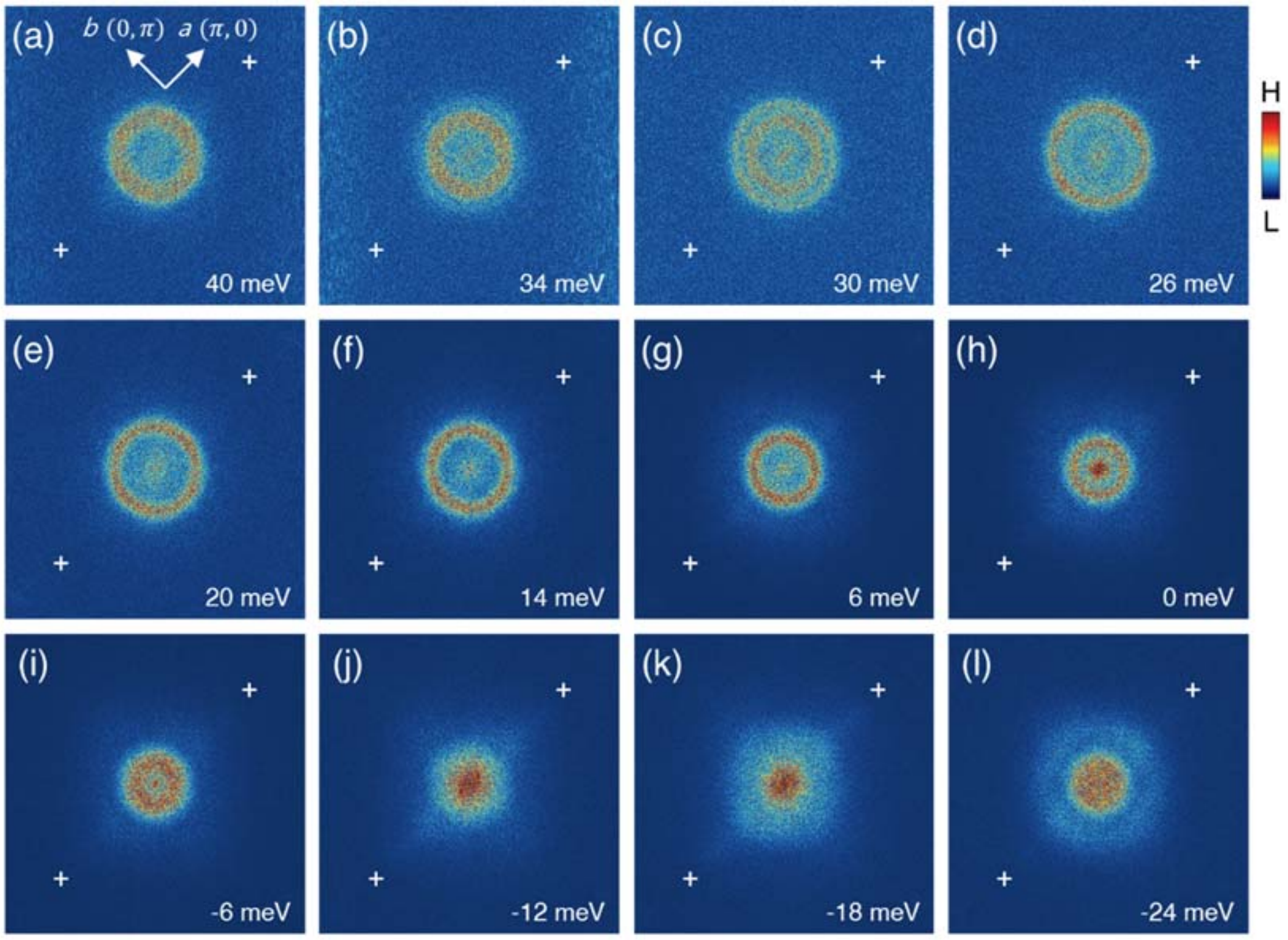

FIG. S13. Raw *dI/dV* data in *q* space corresponding to the data in Fig. S12.

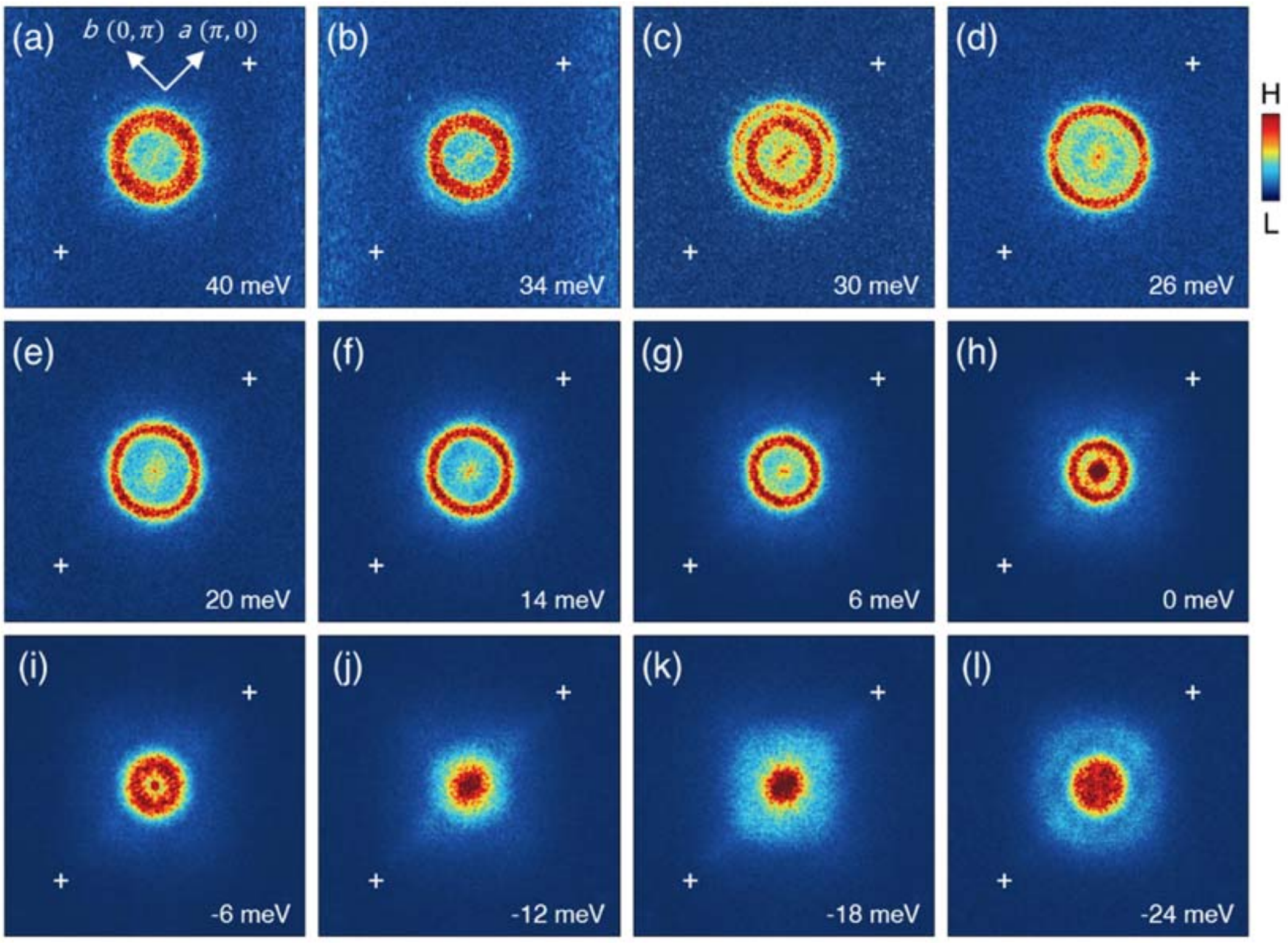

FIG. S14. *dI/dV* data in *q* space, Gaussian averaged with 2 pixel radius, corresponding to the data in Figs. S12 and S13.

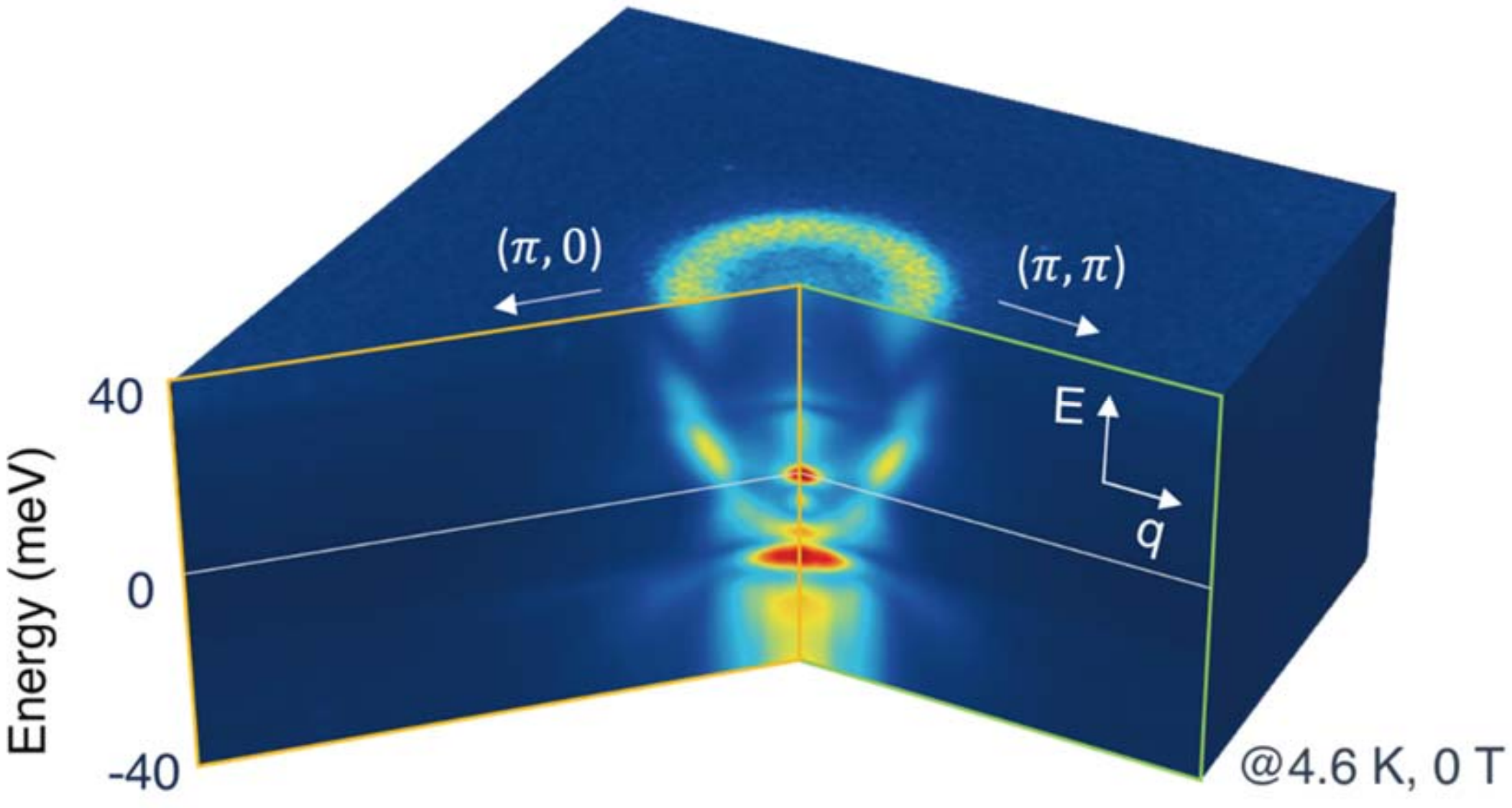

FIG. S15. Relationship of the cross-sectional data in *q*-vs-*E* space and the individual *q* space *dI/dV* data. The top layer shows *q* space image at 40 meV. The two side images show cross-sections in (π, 0) and (π, π) directions after individual *q* space layers are Gaussian-filtered with 3-pixel radius.

## 10. Insensitivity of electronic structure measured with scanning tunneling spectroscopy from details of surface reconstruction

Due to the degeneracy of the ground state surface reconstruction (SR) configurations, the detailed SR patterns can be altered by tunneling current with 300 mV or higher bias voltage as shown in Figs. S16(a)-(c) [34]. Even with the significant change in the SR, the electronic structure measured as tunneling spectra are qualitatively unchanged as shown in Fig. S16(d)-(f) where the slight uneven amplitude changes in (f) may simply be due to the changes in the tunneling matrix elements associated with the significant change in the SR. From this we conclude that the superconducting gap and the higher energy electronic structures we observe in Fig. S16(f) are due to the physics in the FeAs layer and independent of the details of the SR in the topmost SrO layer.

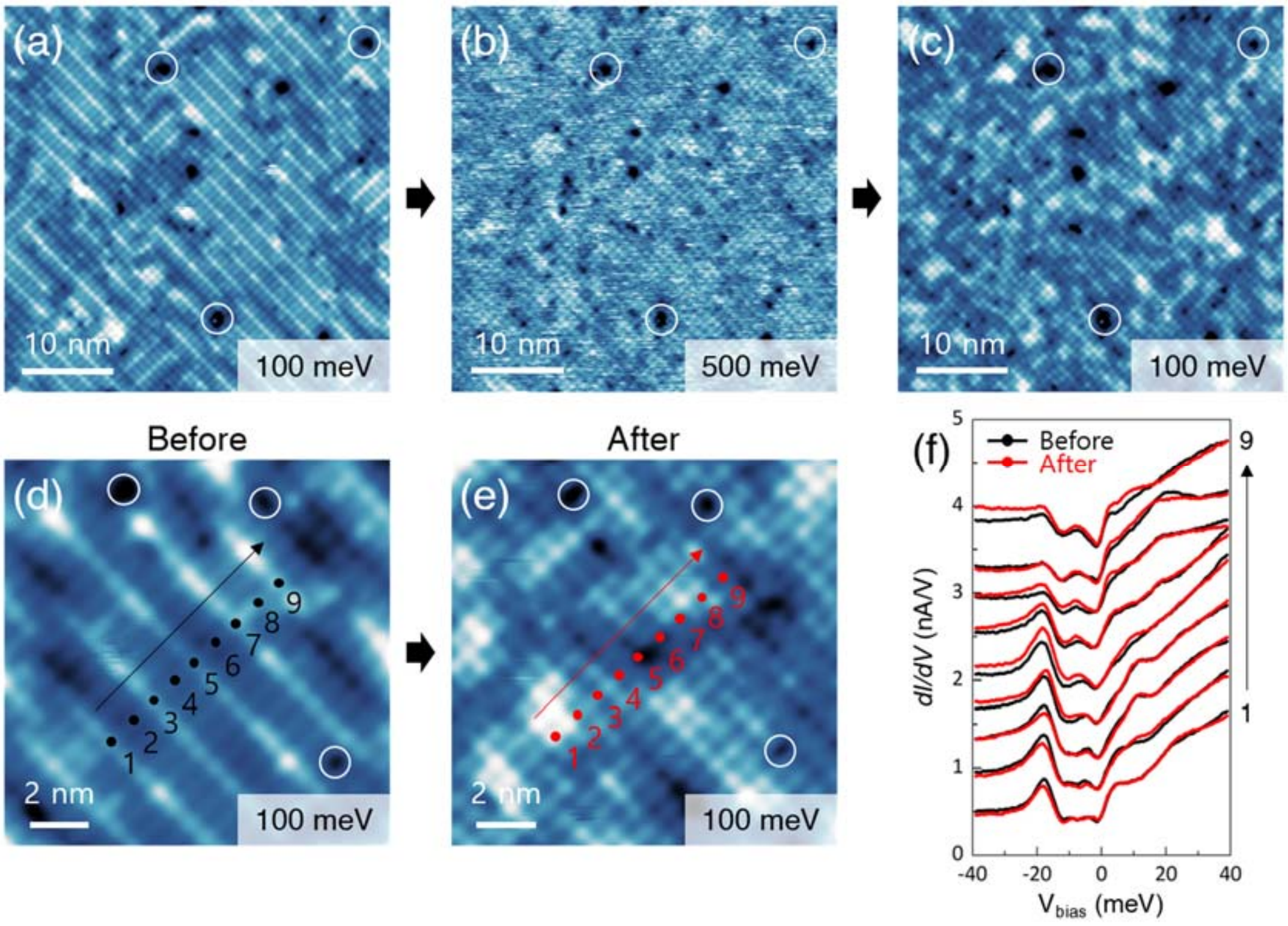

FIG. S16. (a)-(c) As the bias voltage is increased beyond 300 meV, the tunneling current from the W tip significantly modifies the surface reconstruction (SR) of the topmost SrO layer of the cold-cleaved $Sr_2VO_3FeAs$. (d)-(e) High resolution images before and after the SR

modification. (f) The tunneling spectra on the same tip positions before (black) and after (red) the SR modification, where the numbers indicate the corresponding positions of the tip marked in (d) and (e).

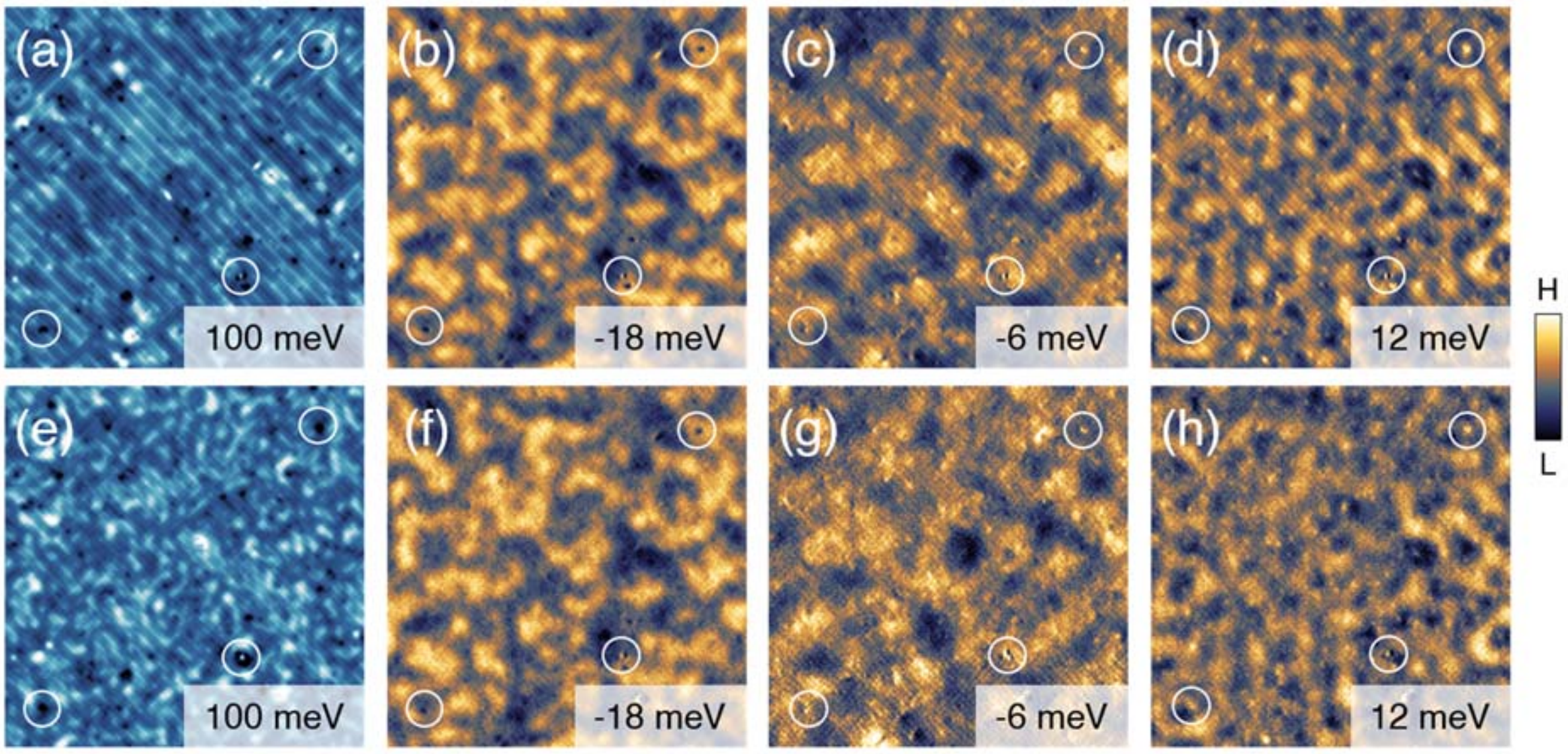

FIG. S17. Electronic structure before and after SR modification. (a) Topograph of as-cleaved $Sr_2VO_3FeAs$. (b)-(d) *dI/dV* images over the same region at three representative energies. (e) Topograph over the same region as in (a) after the SR is modified with high bias scanning. (f)-(h) *dI/dV* images over the same region after the SR modification. The white circles mark the identical locations for all the image panels.

## 11. Electronic structure near the $C_2$ defects

In terms of lattice symmetry and shape, our $C_2$ defects resemble Fe site defects shown in Refs. [35] and [47]. But these Fe site in-plane defects show unitary in-gap states as seen in the above papers which are absent for our $C_2$ defects as shown in Fig. S18 below. The appearance of unitary in-gap states in case of in-plane defects are also observed in the cuprate superconductor by replacing Cu with Ni or Zn impurity atoms [48]. It is reasonable to think that this absence of unitary in-gap state could be the evidence that our $C_2$ defects are off the FeAs plane and the only such possibility considering its symmetry is O vacancy in the $VO_2$ layer.

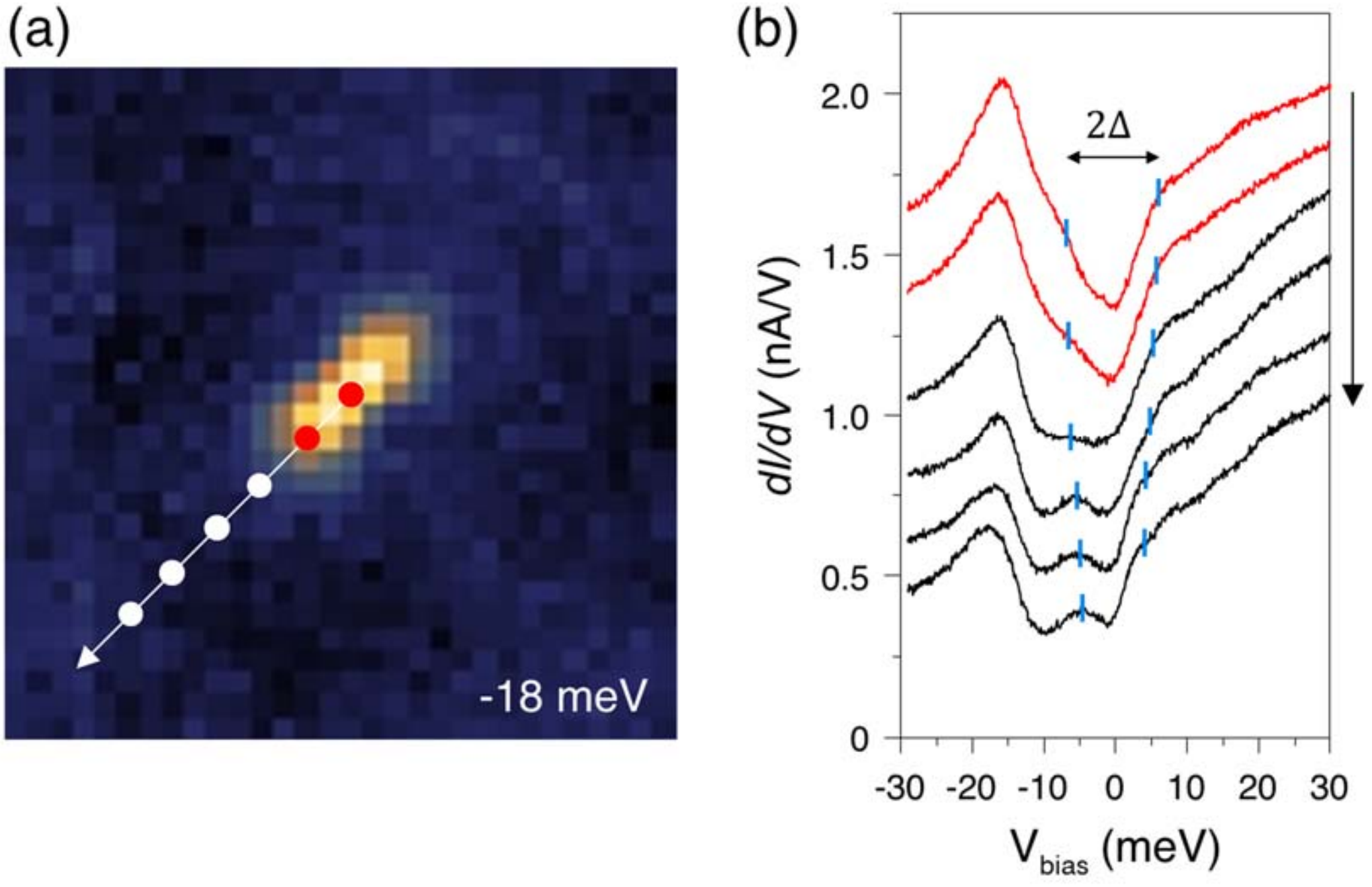

FIG. S18. (a) *dI/dV* image at -18 meV near a $C_2$ defect. (b) *dI/dV* spectra taken at the six locations along the line shown in (a). The blue markers indicate the superconducting coherence peaks extracted by the method used for Figs. 4(b)-(c) and explained in SM section 8.